\documentclass[journal]{IEEEtran}
\pdfoutput=1

\usepackage[ascii]{inputenc}
\usepackage[T1]{fontenc}
\usepackage[final,bookmarksnumbered]{hyperref}
\usepackage[USenglish]{babel}
\usepackage{amsfonts,amsmath,amssymb,mathtools,braket,bbm}
\usepackage[noadjust]{cite}
\usepackage{textcase,times,microtype,url}
\usepackage{aliascnt,graphicx,color}

\newcommand{\newjointcountertheorem}[3]{\newaliascnt{#1}{#2}\newtheorem{#1}[#1]{#3}\aliascntresetthe{#1}}
\newtheorem{thm}{Theorem}
\newjointcountertheorem{lem}{thm}{Lemma}
\newjointcountertheorem{prp}{thm}{Proposition}
\newjointcountertheorem{cor}{thm}{Corollary}
\newjointcountertheorem{exl}{thm}{Example}

\def\Snospace~{\S{}}
\makeatletter
\def\p@subsection{\thesection.}
\makeatother

\DeclareMathOperator{\diag}{diag}
\DeclareMathOperator{\Sym}{Sym}

\DeclareMathOperator{\spec}{spec}

\DeclareMathOperator{\tr}{tr}

\DeclareMathOperator{\supp}{supp}
\DeclareMathOperator{\SU}{SU}

\renewcommand{\epsilon}{\varepsilon}

\newcommand{\norm}[1]{\lVert#1\rVert}
\newcommand{\abs}[1]{\lvert#1\rvert}
\newcommand{\PP}{\mathbb P}
\newcommand{\CC}{\mathbb C}
\newcommand{\RR}{\mathbb R}

\newcommand{\ZZ}{\mathbb Z}

\newcommand{\id}{\mathbbm 1}

\newcommand{\proj}[1]{\lvert#1\rangle\langle#1\rvert}

\allowdisplaybreaks[4]

\begin{document}

\title{Lower Bounds for Quantum Parameter Estimation}
\author{Michael~Walter and Joseph~M.~Renes,~\IEEEmembership{Member,~IEEE}%
\thanks{This work is supported by the European Research Council (grant 258932), the German Science Foundation (grant CH 843/2--1),
the Swiss National Science Foundation (grants 200020\_135048, PP00P2\_128455, and 20CH21\_138799),
and the National Center of Competence in Research `Quantum Science and Technology'.
M.~Walter was also supported by the Isaac Newton Institute for Mathematical Sciences during a research stay in fall 2013, where part of this work was completed. %
Parts of this work were presented at the 2014 IEEE International Symposium on Information Theory.
}%
\thanks{M.~Walter was with the Institute for Theoretical Physics, ETH Zurich, 8093 Z\"urich, Switzerland. He is now with the Stanford Institute for Theoretical Physics, Stanford University, Stanford, CA 94305 (e-mail: \href{mailto:michael.walter@stanford.edu}{michael.walter@stanford.edu}).}%
\thanks{J.~M.~Renes is with the Institute for Theoretical Physics, ETH Zurich, 8093 Z\"urich, Switzerland (e-mail: \href{mailto:renes@phys.ethz.ch}{renes@phys.ethz.ch}).}%
\thanks{Copyright (c) 2014 IEEE. Personal use of this material is permitted.  However, permission to use this material for any other purposes must be obtained from the IEEE by sending a request to pubs-permissions@ieee.org.}}
\maketitle

\begin{abstract}
The laws of quantum mechanics place fundamental limits on the accuracy of measurements and therefore on the estimation of unknown parameters of a quantum system.
In this work, we prove lower bounds on the size of confidence regions reported by any region estimator for a given ensemble of probe states and probability of success.
Our bounds are derived from a previously unnoticed connection between the size of confidence regions and the error probabilities of a corresponding binary hypothesis test.
In group-covariant scenarios, we find that there is an ultimate bound for any estimation scheme which depends only on the representation-theoretic data of the probe system, and we evaluate its asymptotics in the limit of many systems, establishing a general ``Heisenberg limit'' for region estimation.
We apply our results to several examples, in particular to phase estimation, where our bounds allow us to recover the well-known Heisenberg and shot-noise scaling.
\end{abstract}

\begin{IEEEkeywords}
Quantum parameter estimation, confidence regions, lower bounds, Heisenberg limit, covariant estimation, hypothesis testing, quantum information theory, representation theory.
\end{IEEEkeywords}

{\small The results presented in this article have been announced in \cite{announcement}.}

\section{Introduction and Summary of Results}

The estimation of unknown parameters in a quantum system is a basic problem in quantum mechanics \cite{helstrom_1974_estimation,holevo_1978_estimation,werner_1985_screen} and its many applications.
For example, in the Cesium atomic clock, the unknown frequency shift of a quartz oscillator is to be estimated by measuring Cesium atoms that have been subjected to a time evolution that depends on the frequency shift, while in a Mach--Zehnder interferometer one tries to estimate the phase shift between two optical paths from the output beams of photons.
The standard way to reduce the uncertainty inherent with any measurement, quantum or otherwise, is to repeat the experiment a large number of times.
Equivalently, one may take many probes in an i.i.d.\ state $\rho^{\otimes N}$ and subject them all to the same dynamics.
The mean square error of such a scheme scales as $1/N$, as one would expect from the central limit theorem (even when one is allowed to perform entangled measurements); this is known as the standard quantum limit or \emph{shot noise limit} \cite{giovannettilloydmaccone06}.
Interestingly, it is often possible to achieve better precision by using entangled probe states \cite{giovannettilloydmaccone04}. Here, it can be established by using the quantum Cram\'{e}r--Rao inequality \cite{helstrom_1974_estimation,braunstein_caves_1994_statistical} that the ultimate lower bound on the mean-square error for any unbiased estimator, known as the \emph{Heisenberg limit}, scales as $1/N^2$ \cite{giovannettilloydmaccone06}. There exist families of probe states that achieve this scaling even when no prior information about the parameter is available \cite{summy_pegg_1990_phase,sandersmilburn95,berrywiseman00}.

Recently, it has become of interest in quantum estimation theory to consider \emph{region estimators}, i.e.\ estimators that report a confidence region rather than a single point in the parameter space \cite{neyman_1935_problem}. This is a way of providing rigorous error bars, and it avoids conceptual problems that are inherent with point estimators. See e.g.\ \cite{KahnGuta09,gross_liu_flammia_et_al_2010_quantum,blumekohout10b,blumekohout_2012_robust,christandl_renner_2012_reliable,YamagataFujiwaraGill13} for results in the context of quantum state tomography \cite{paris_rehacek_2004_quantum}, which is perhaps the ultimate parameter estimation problem, as the entire quantum state is unknown.
There is in fact a close connection between mean-square error and confidence regions:
By Chebyshev's inequality,
any unbiased point estimator with mean-square error $\Delta^2$ can be considered as a region estimator with success probability $p_\text{succ}$ by reporting an interval of radius $\delta = \Delta/\sqrt{1-p_\text{succ}}$ around its estimate.

\medskip

In this work, we are interested in the fundamental limits of any region estimation scheme. More precisely, our goal is to prove lower bounds on the maximal volume $V_{\max}$ and average volume $V_{\text{avg}}$ of the confidence regions reported by an arbitrary region estimator, depending only on the success probability $p_{\text{succ}}$ of the estimator and the ensemble of probe states $\{p_X^x, \rho^x_B\}$, where $X$ is the parameter space and $B$ the probe system (see \autoref{sec:region estimators} for precise definitions). We work in the Bayesian scenario, motived by the information-theoretic methods that we employ to prove our bounds. But any Bayesian lower bound implies directly a lower bound for the minimax, i.e.\ worst-case performance: If a region estimator has success probability $p_\text{succ}$ for any fixed value of the unknown parameter, then it also succeeds with probability at least $p_\text{succ}$ for an arbitrary prior, and hence our bounds apply.
Furthermore, it is often natural to consider a prior distribution over the parameter (see e.g.\ \cite{fraas13}, where a Bayesian variant of the Cram\'{e}r--Rao bound has been used to study the steady state of atomic clocks), and doing so also allows comparison of the local and global performance of estimation schemes (see e.g.\ \cite{durkin_dowling_2007,hall_wiseman_2012} and below).

Our starting point is the duality between region estimators and hypothesis tests, well-known from statistics \cite[\S 3.5]{lehmannromano05}. From a given region estimator, we construct the binary hypothesis test on the bipartite classical--quantum system $XB$ that rejects the null hypothesis unless the state of the classical system $X$ is contained in the region predicted by the estimator from the probe system $B$. If we take as null hypothesis the cq-state $\rho_{XB}$ that corresponds to the ensemble of probe states, then the probability that the test correctly identifies this null hypothesis is precisely equal to the success probability $p_\text{succ}$ of the original region estimator (i.e., the type I-error is $1-p_\text{succ}$).

As the alternative hypothesis, consider any ``uncoupled'' quantum state of the form $\id_X/\abs X \otimes \sigma_B$.
For fixed values of the parameter $x$, the estimator now reports a region independent of $x$, and because the parameter values are chosen uniformly at random, the probability that the test wrongly rejects this alternative hypothesis is related to the size of the region.
Specifically, we show that the type II-error is never larger than $V_{\max} / \abs X$, the maximal relative volume reported by the region estimator from which we constructed the test.
By minimizing the type II-error over all such tests (in particular, over all estimators for the given probability of success) and maximizing over all choices of $\sigma_B$, we thus obtain the following \emph{hypothesis-testing lower bound}, which holds for an arbitrary region estimator (\autoref{hypothesis-testing lower bound} in \autoref{sec:region estimators}):
\begin{equation}
\label{intro:hypothesis-testing lower bound}
  \frac {V_{\max}} {\abs X} \geq \sup_{\sigma_B} \beta_{p_\text{succ}}(\rho_{XB}, \frac {\id_X} {\abs X} \otimes \sigma_B).
\end{equation}
Here and in the following, $\beta_\alpha(\rho, \sigma)$ denotes the minimal type-II error for any hypothesis test with null hypothesis $\rho$ and alternative hypothesis $\sigma$, if we require that the null hypothesis is correctly identified with probability $\alpha$ or larger.
The bound \eqref{intro:hypothesis-testing lower bound} is completely independent of the inner workings of the region estimators -- it depends only on the ensemble of probe states and on the desired success probability. To our knowledge, this connection between the volume of the confidence regions and the type-II error in binary hypothesis testing has not been noticed before \footnote{Interestingly, the family of Ziv--Zakai inequalities \cite{ziv_zakai_1969_lower,tsang_2012_zivzakai} give lower bounds on the mean square error of point estimators in terms of \emph{Bayesian} binary hypothesis testing, with alternative hypotheses the shifted probe states $\rho^{x+\tau}_B$ for varying choices of $\tau$.}.

By choosing $\sigma_B$ as the average probe state $\int dx \, p^x_X \rho^x_B$ instead of optimizing over all $\sigma_B$ in \eqref{intro:hypothesis-testing lower bound}, we obtain a lower bound for the average volume $V_\text{avg}$.
Similarly, by optimizing over all $\sigma_B = \rho^x_B$, we obtain a lower bound for the average volume for any fixed value of the parameter, $\sup_x V_\text{avg}(x)$ \cite{blumekohout_2012_robust}.
We also describe a general procedure for deducing lower bounds on the average volume directly from \eqref{intro:hypothesis-testing lower bound} and the results presented below (see discussion after \autoref{hypothesis-testing lower bound}).

The right-hand side of \eqref{intro:hypothesis-testing lower bound} is closely connected to the \emph{conditional hypothesis-testing entropy}, which is defined as $\log \sup_{\sigma_B} \beta_\alpha(\rho_{XB}, \id_X \otimes \sigma_B)$ and which shares many formal properties with the conditional von Neumann entropy \cite{wangrenner12,hayashitomamichel12,generalizedentropies12}. In this way the lower bound \eqref{intro:hypothesis-testing lower bound} acquires an intuitive information-theoretical interpretation:
\emph{The maximal log-volume reported by the estimator is at least as large as the conditional hypothesis-testing entropy of the parameter $X$ conditioned on the probe system $B$}.
The larger the uncertainty in the parameter given the probe system, as measured by the conditional entropy, the larger the maximal region estimator volume.
The lower bound can also be understood as the quantum-mechanical variant of a converse bound for joint source-channel coding in \cite{kostinaverdu13}, adapted to the case of trivial encoder. Indeed, the connection between information theory and statistics has a long and fruitful history.
In the context of quantum parameter estimation, lower bounds to the mean-square error in terms of the conditional von Neumann entropy have been derived for the first time in \cite{yuen_1992} by using rate-distortion theory (cf.\ the references in \cite{hall_wiseman_2012}). Entropic lower bounds can also be established as a consequence of \eqref{intro:hypothesis-testing lower bound}, both for region and for point estimation (\autoref{entropic lower bound} and \autoref{mse entropic}).

We then focus on \emph{group-covariant} scenarios, where the family of probe states is obtained from an initial state $\rho^{x_0}_B$ by the action of a compact Lie group $G$, $\rho^{g x_0}_B = g \rho^{x_0}_B g^{-1}$ \cite{helstrom_1974_estimation,holevo_1978_estimation,holevo82,werner_1985_screen,chiribella_dariano_perinotti_et_al_2004_covariant,chiribella_dariano_perinotti_et_al_2004_efficient,chiribella_dariano_sacchi_2005_optimal,hayashi_2011_comparison,hayashi_2013_fourier} (\autoref{sec:covariant}).
Mathematically, $X$ is a homogeneous space.
Here, \eqref{intro:hypothesis-testing lower bound} can be ``untwisted'' (\autoref{covariant lower bound});
in particular, for the uniform prior we obtain that
\begin{equation}
\label{intro:covariant lower bound}
  \frac {V_{\max}} {\abs X} \geq \sup_{\tilde\sigma_B} \beta_{p_\text{succ}}(\rho_B^{x_0}, \tilde\sigma_B),
\end{equation}
where the optimization is now over \emph{$G$-invariant} states $\tilde\sigma_B$.
The right-hand side of \eqref{intro:covariant lower bound} is a one-shot analog of the \emph{$G$-asymmetry} or \emph{relative entropy of frameness} \cite{vaccaro_et_al_2008,gourmarvianspekkens09} (cf.\ \cite{schuchverstraetecirac04}).
By minimizing over all probe states $\rho_B^{x_0}$, we establish a fundamental lower bound that holds for arbitrary region estimators and probe states, depending only on the representation-theoretic data of the probe-system Hilbert space $\mathcal H_B$ (\autoref{state-independent covariant lower bound}). In the case where $X = G$, it takes the particularly simple form
\begin{equation}
\label{intro:covariant state-independent lower bound}
  \frac {V_{\max}} {\abs X} \geq \frac {\beta_{p_\text{succ}}(p_X, {\id_X} / {\abs X})} {\sum_\lambda d_\lambda r_\lambda}.
\end{equation}
Here, $\lambda$ labels the irreducible representations of $G$ that occur in $\mathcal H_B$, $d_\lambda$ denotes the corresponding dimension, and $r_\lambda := \min \{ d_\lambda, m_\lambda \}$, with $m_\lambda$ the multiplicity.
Observe that \eqref{intro:covariant state-independent lower bound} is essentially independent of the multiplicities $m_\lambda$ (small multiplicities can improve the bound, but large multiplicities do not enter). This can be rather intuitively understood: Since the group $G$ acts the same way on each copy of an irreducible representation, the multiplicities should only matter insofar as they allow for entanglement between the representation and the multiplicity space, and this entanglement is bounded by $r_\lambda \leq d_\lambda$ \cite{chiribella_dariano_sacchi_2005_optimal}.
The numerator in \eqref{intro:covariant state-independent lower bound} measures the deviation of the prior from being uniform; it is equal to $p_\text{succ}$ if the prior is uniform, and otherwise smaller.

We now consider the asymptotics for $N \rightarrow \infty$ copies of the probe system (\autoref{sec:asymptotics}).
Here, we establish a lower bound of the order $1/N^{\dim G}$ that holds for arbitrary $X = G$ (\autoref{state-independent covariant heisenberg bound}). This is a very general ``Heisenberg limit'' for the volume of confidence regions. As a direct consequence, we obtain lower bounds on the mean-square error of point estimators, thereby generalizing results in the literature for $U(1)$ \cite{giovannettilloydmaccone06} and $\SU(d)$ \cite{chiribella_dariano_sacchi_2005_optimal,kahn_2007_fast}.

Phase estimation is an important instance of covariant estimation. Here, the probe states are obtained by the evolution of an initial state under a given periodic Hamiltonian, $\rho^\theta_B = e^{\imath \theta H} \rho^0_B e^{-\imath \theta H}$, with $\theta \in [0,2\pi]$ the unknown phase, and the lower bound \eqref{intro:covariant state-independent lower bound} reduces to
\begin{equation}
  \label{intro:phase-estimation lower bound}
  \frac {V_{\max}} {2\pi} \geq \frac {\beta_{p_\text{succ}}(p_X, {\id_X} / {2 \pi})} J,
\end{equation}
where $J$ denotes the number of distinct eigenvalues of the Hamiltonian $H$.
In particular, consider a single-body Hamiltonian of the form $H_N = \sum_n H^{(n)}$ acting on $N$ particles, with each $H^{(n)}$ acting in the same way.
It is easy to see that the number of eigenvalues of any such periodic Hamiltonian scales at most linearly with $N$, so that $J = O(N)$ in \eqref{intro:phase-estimation lower bound}.
Heisenberg scaling for the mean-square error of point estimators is an easy consequence (\autoref{subsec:phase-estimation}). Conversely, since the latter is well-known to be achievable \cite{summy_pegg_1990_phase,sandersmilburn95,berrywiseman00}, there exist confidence regions for phase estimation whose volume scales as $1/N$. Thus our lower bound is necessarily tight (up to constants).
In contrast, \emph{separable} probe states fulfill a tighter lower bound of order $1/\sqrt N$. This is because any pure separable probe state is concentrated only on $O(\sqrt N)$ many eigenvectors of the Hamiltonian.
%
More generally, \eqref{intro:phase-estimation lower bound} implies that the global precision in phase estimation can only be improved by either acquiring additional prior information or increasing the number of eigenvalues of the Hamiltonian. This is in stark contrast to local arguments based on the Cram\'{e}r--Rao bound, which show that the local sensitivity of an estimator is determined by the gap between the minimal and maximal eigenvalue; see \cite{durkin_dowling_2007,hall_wiseman_2012} for discussions in the context of the mean-square error.
This phenomenon is also visible on the level of individual probe states. For example, the oft-mentioned GHZ or NOON state $( \ket 0^{\otimes N} + \ket 1^{\otimes N} ) /{\sqrt 2}$ is supported only on two eigenspaces of the Hamiltonian $H = \sum_n \sigma_z^{(n)}$, and hence is only useful if prior information on the phase is available; any ostensible scaling of the precision with $N$ can be seen purely a consequence of the prior information available about the phase \cite{hall_wiseman_2012}. This can be directly verified by evaluating \eqref{intro:covariant lower bound} for the GHZ state (\autoref{phase estimation numerics figure}).
For a different instance of this distinction between local and global performance, we also consider a ``non-linear'' Hamiltonian, where $N$ probes are coupled via two-body interactions to an auxiliary system, $H_{N+1} = \sum_{n=1}^N\sigma_z^{(n)} \sigma_z^{(N+1)}$ \cite{braun_martin_2011_heisenberglimited,braun_popescu_2013_coherently}.
This interaction can generate entanglement, and local arguments suggest that $1/N$ scaling should be possible even if the initial probe state $\rho^{x_0}_B$ is separable. By evaluating \eqref{intro:covariant lower bound}, we show that this conclusion does not hold globally; instead, we find a lower bound of order $1/\sqrt N$.
We also revisit some other paradigmatic scenarios, such as phase estimation with energy-bounded probe states and state estimation of a pure state, and establish corresponding lower bounds (\autoref{sec:example}).

\paragraph*{Notation and Conventions} We denote the set of states, i.e.\ density operators, on a Hilbert space $\mathcal H_B$ by $\mathcal S_B$, and the set of observables, i.e.\ Hermitian operators, by $\mathcal O_B$. The expectation value of an observable $O_B$ in state $\rho_B$ will be denoted by the pairing $\braket{\rho_B, O_B} := \tr \rho_B O_B$.
A POVM with outcomes in a measurable space $Y$ is a $\sigma$-additive function $M_B$ that assigns to each measurable subset $O \subseteq Y$ a positive semidefinite operator $M_B(O)$, with $M_B(\emptyset) = 0$ and $M_B(Y) = \id_B$.
For any state $\rho_B$, $\braket{\rho_B, M_B(-)}$ is the probability distribution of measurement outcomes.
We will later generalize these definitions to bipartite classical--quantum systems with continuous classical parameter $X$.
We shall often write $f^x$ for the value of a function $f$ at some point $x$, and $f = (f^x)$ for the function itself (which is a common notation when $X$ is finite or discrete).
Throughout this article, we will use subscripts to indicate systems, such as the classical system $X$ and the quantum system $B$.

\section{Region Estimators and the Hypothesis-Testing Lower Bound}
\label{sec:region estimators}

Let $X$ be the measurable space of \emph{parameters}, equipped with a measure $\mu_X$, which we shall later use to measure the volume of confidence regions, such that $(X,\mu_X)$ is a standard measure space. Let $p_X = (p_X^x) \in L^1(X, \mu_X)$ be the probability density of the \emph{prior distribution}. For each parameter value $x \in X$ we are given a \emph{probe state} $\rho^x_B$ on some (possibly infinite-dimensional) Hilbert space $\mathcal H_B$, and we shall assume that the family $(\rho^x_B)$ is $\mu_X$-measurable in $x$ (see Appendix~\ref{measurability appendix} for a precise definition).
The presented theory can be generalized readily to the case where the probe states are normal states on a general von Neumann algebra, but we will not need this here.

\begin{figure}
\begin{center}
\includegraphics[width=\linewidth]{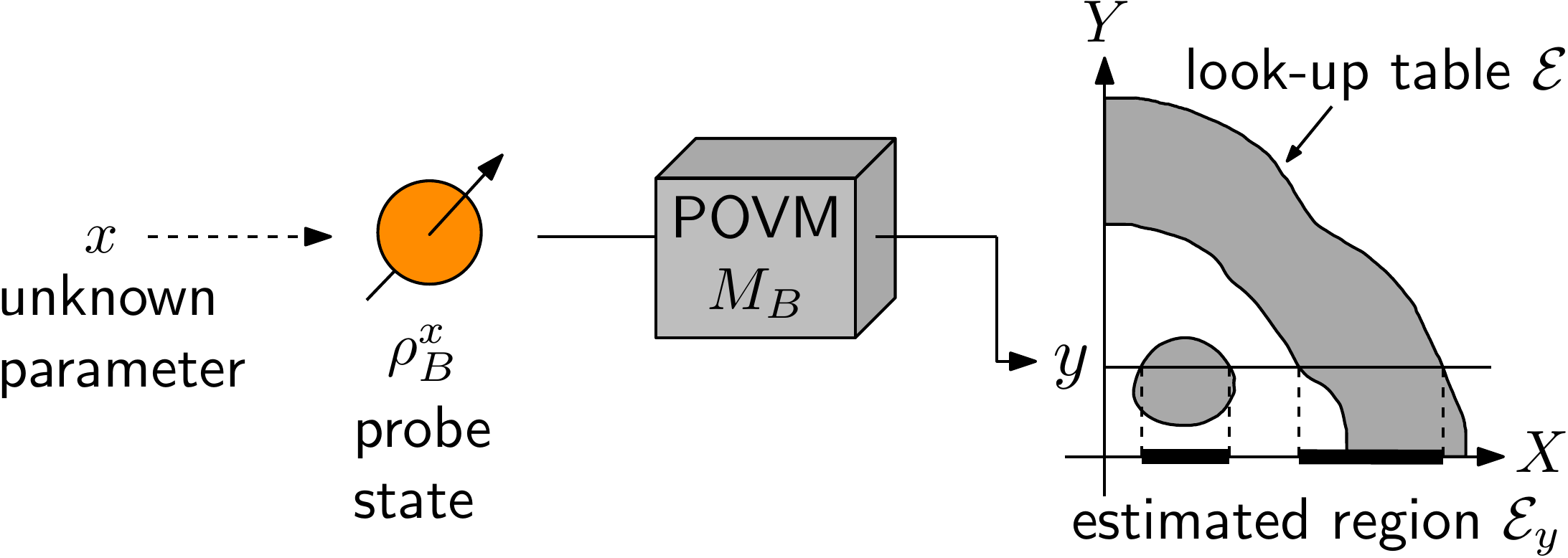}
\end{center}
\caption{\emph{Definition of a region estimator.} A POVM measurement $M_B$ is performed on the probe system. The region estimate is then obtained from the look-up table $\mathcal E$ as the set of all $x$ that are compatible with the measurement outcome $y$.}
\label{region estimator figure}
\end{figure}

A \emph{region estimator} for this scenario consists of a POVM $M_B$ on $\mathcal H_B$ with outcomes in some measurable space $Y$, together with a measurable subset $\mathcal E \subseteq X \times Y$, the \emph{look-up table}. The interpretation is that the experimenter first performs the POVM measurement and obtains some result $y \in Y$ according to the probability measure $\braket{\rho, M_B(-)}$; the estimated region is then given by the set
$\mathcal E_y := \{ x : (x,y) \in \mathcal E \}$ (\autoref{region estimator figure}). In other words, if $x$ is the true value of the underlying parameter then the estimator succeeds if the POVM measurement outcome is an element of the ``compatible set'' $\mathcal E_x = \{ y : (x,y) \in \mathcal E \}$.
Thus the \emph{(average) success probability} of the estimator is
\begin{equation}
  \label{eq:success probability}
  p_{\text{succ}} = \int_X d\mu_X(x) \, p^x_X \braket{\rho^x_B, M_B(\mathcal E_x)},
\end{equation}
and the \emph{maximal volume} of any region reported by the estimator is
\begin{equation}
  \label{eq:maximal volume}
  V_{\max} = \sup_y \mu_X(\mathcal E_y) \in [0,\infty].
\end{equation}
We show in Appendix~\ref{measurability appendix} that our assumptions guarantee that these quantities are in fact well-defined.

The above way of encoding a random subset using a ``look-up table'' avoids subtle measurability issues that may arise if one tries to define random subsets as subset-valued random variables \cite{tsirelson_lectures}.
The common case where the reported regions are $\delta$-balls with respect to some metric $d_X$ is recovered by setting $Y = X$ and $\mathcal E = \{ (x, y) : d_X(x, y) \leq \delta \}$.

\begin{lem}
  For any region estimator $(M_B, \mathcal E)$ and any quantum state $\sigma_B$ on $\mathcal H_B$, we have the lower bound
  \begin{equation}
    \label{eq:max volume bound}
    V_{\max} \geq \int_X d\mu_X(x) \braket{\sigma_B, M_B(\mathcal E_x)}.
  \end{equation}
  In fact, the right-hand side is equal to the average volume of the region estimate in state $\sigma_B$.
\end{lem}
\begin{IEEEproof}
  Let us denote by $\nu_Y = \braket{\sigma_B, M_B(-)}$ the probability distribution of POVM measurement outcomes in the state $\sigma_B$.
  Then
  \[ \int_Y d\nu_Y(y) \, \mu_X(\mathcal E_y) \leq V_{\max}, \]
  since the left-hand side is the average volume of the corresponding region estimate.
  On the other hand,
  \begin{align*}
    &\quad \int_Y d\nu_Y(y) \, \mu_X(\mathcal E_y) \\
    &= \int_Y d\nu_Y(y) \int_X d\mu_X(x) \, \id_{\mathcal E}(x, y) \\
    &= \int_X d\mu_X(x) \int_Y d\nu_Y(y) \, \id_{\mathcal E}(x, y) \\
    &= \int_X d\mu_X(x) \braket{\sigma_B, M_B(\mathcal E_x)}
  \end{align*}
  where we have used Tonelli's theorem to swap the order of integration in the second step \cite[Theorem 2.36]{folland99}.
\end{IEEEproof}

Interestingly, both the definition of the success probability \eqref{eq:success probability} and the lower bound \eqref{eq:max volume bound} depend only on the ensemble of probe states $\{ p^x_X, \rho^x_B \}$ and on the operators $E^x_B := M_B(\mathcal E_x)$.
Now, $(E^x_B)$ is an operator-valued function of $x$ which takes values between $0$ and $\id_B$, and so can be interpreted as a \emph{POVM element} on the classical--quantum system $XB$ (see below for a precise definition). For this POVM element, \eqref{eq:success probability} shows that
$\int_X d\mu_X(x) \, p^x_X \braket{\rho^x_B, E^x_B} \geq p_{\text{succ}}$
holds by definition (even with equality), and the lower bound \eqref{eq:max volume bound} asserts that
$V_{\max} \geq \int_X d\mu_X(x) \braket{\sigma_B, E^x_B}.$
Therefore, if we minimize the right-hand side of the latter over all POVM elements $(E^x_B)$ that satisfy the former condition then we obtain a universal relation between success probability and maximal volume that holds for \emph{all} region estimators for a given ensemble $\{p^x_X, \rho^x_B\}$:
We find that
\begin{equation}
\label{eq:hands-on hypothesis testing}
\begin{aligned}
  V_{\max}
  \geq
  \inf
  \{
  &\int_X d\mu_X(x) \braket{\sigma_B, E^x_B}
  : 0 \leq E^x_B \leq \id_B, \\
  &\int_X d\mu_X(x) \, p^x_X \braket{\rho^x_B, E^x_B} \geq p_{\text{succ}}
  \}.
\end{aligned}
\end{equation}
for all states $\sigma_B$.
We note that the integral of the 
function $\braket{\sigma_B, E^x_B}$ can be infinite if $X$ has infinite measure (e.g., when estimating a shift parameter in $\RR$). This makes sense, since in this case an estimator might report regions of infinite volume. However, even if $X$ has infinite measure the lower-bound is still mathematically correct and can be finite.

If, on the other hand, we assume that $X$ has finite measure then \eqref{eq:hands-on hypothesis testing} has a direct interpretation in terms of binary hypothesis testing. Making this precise is slightly complicated by the fact that, formally, the ensemble $\{p^x_X, \rho^x_B\}$ cannot be described as a density operator if $X$ is a continuous space -- indeed, the usual identification between ensembles $\{p^x_X, \rho^x_B\}$ and density operators $\sum_x p^x_X \proj x \otimes \rho^x_B$ is only possible for discrete $X$.

\medskip\subsubsection*{Classical--Quantum Hypothesis Testing}
Instead, we should model a bipartite classical--quantum system with classical part $(X,\mu_X)$ and quantum part $\mathcal H_B$ by the von Neumann algebra $\mathcal M_{XB} = L^\infty(X, \mu_X; B(\mathcal H_B))$. Its elements are operator-valued functions $E_{XB} = (E^x_B) \colon X \rightarrow B(\mathcal H_B)$ with finite norm $\norm{E_{XB}}_\infty = \sup_x \norm{E^x_B}_\infty$.
Thus a \emph{classical--quantum observable (cq-observable)} is a (weak-$\star$-measurable) bounded function on $X$ that takes values in the space of Hermitian operators on $\mathcal H_B$, i.e.\ an element of
\begin{equation*}
  \mathcal O_{XB} = \{ E_{XB} = (E^x_B) : \norm{E_{XB}}_\infty < \infty, \, E^x_B \in \mathcal O_B \},
\end{equation*}
Similarly, a \emph{cq-state} is a ($\mu_X$-measurable) normalized function on $X$ that takes values in the space of positive semidefinite operators on $\mathcal H_B$, i.e.\ an element of
\begin{equation*}
  \mathcal S_{XB} = \{ \rho_{XB} = (\rho^x_B) : \norm{\rho_{XB}}_1 = 1, \, \rho^x_B \geq 0 \},
\end{equation*}
where $\norm{f_{XB}}_1 := \int d\mu_X(x) \norm{f^x_B}_1$. Any positive multiple of such a state is called an \emph{unnormalized cq-state}.
It will be convenient to use the notation $p_X \otimes \rho_B := (p^x_X \rho_B)$ for $p_X = (p^x_X) \in L^1(X, \mu_X)$ and $\rho_B$ a density operator on $\mathcal H_B$.
The \emph{expectation value} of an observable $E_{XB} = (E^x_B)$ in the state $\rho_{XB} = (\rho^x_B)$ is given by the pairing
\begin{equation*}
  \braket{\rho_{XB}, E_{XB}} = \int d\mu_X(x) \braket{\rho^x_B, E^x_B}.
\end{equation*}
Finally, a \emph{cq-POVM} with outcomes in some measurable space $Y$ is a function $M_{XB}$ that assigns to each measurable set $O \subseteq Y$ a positive semidefinite element $M_{XB}(O) = (M^x_B(O))$ of $\mathcal O_{XB}$, with $M_{XB}(\emptyset) = 0$ and $M_{XB}(Y) = \id_{XB}$, and which is (weakly) $\sigma$-additive \cite{holevo82}. For each state $\rho_{XB}$, the probability that the measurement outcome is an element of $O \subseteq Y$ is given by the formula
\begin{equation*}
  \braket{\rho_{XB}, M_{XB}(O)} = \int d\mu_X(x) \braket{\rho^x_B, M^x_B(O)}.
\end{equation*}
These definitions arise naturally in the algebraic approach to quantum mechanics \cite{yuenozawa93,berta_christandl_furrer_et_al_2013_continuous}, and they reduce to the usual definitions if $X$ is finite or discrete.

A \emph{(binary) cq-hypothesis test} with null hypothesis $\rho_{XB}$ and alternative hypothesis $\sigma_{XB}$ is a measurement with two possible outcomes, $\rho_{XB}$ or $\sigma_{XB}$. It is fully determined by the POVM element $E_{XB} = M_{XB}(\{\rho_{XB}\})$ corresponding to the null hypothesis.
We set
\begin{equation}
\label{cq hypothesis primal}
\begin{aligned}
  \beta_\alpha(\rho_{XB}, \sigma_{XB}) = \inf \{ \braket{\sigma_{XB}, E_{XB}} :  E_{XB} \in \mathcal O_{XB}, \\
  0 \leq E_{XB} \leq \id_{XB},
  \braket{\rho_{XB}, E_{XB}} \geq \alpha \}.
\end{aligned}
\end{equation}
It is the minimal type-II error if we require that the type-I error be no larger than $1 - \alpha$.

Using this notation, and optimizing over all states $\sigma_B$, we can now rewrite \eqref{eq:hands-on hypothesis testing} in the following way (still assuming that $\abs X := \mu_X(X) < \infty$ so that $\id_X / \abs X \otimes \sigma_B$ is a normalized cq-state):

\begin{thm}[Hypothesis-testing lower bound]
  \label{hypothesis-testing lower bound}
  For any region estimator, we have that
  \begin{equation}
    \label{eq:hypothesis-testing lower bound}
    \frac {V_{\max}} {\abs X} \geq \sup_{\sigma_B} \beta_{p_{\text{succ}}}(\rho_{XB}, \frac {\id_X} {\abs X} \otimes \sigma_B),
  \end{equation}
  where $\rho_{XB} = (p^x_X \rho^x_B)$ is the cq-state corresponding to the ensemble of probe states $\{p^x_X,\rho^x_B\}$.
\end{thm}

\autoref{hypothesis-testing lower bound} allows us to lower-bound the maximal volume reported by an arbitrary region estimator by analyzing binary hypothesis tests between the ensemble of probe states and arbitrary ``decoupled'' probe states of the form $\id_X / \abs X \otimes \sigma_B$.

\medskip\subsubsection*{Average vs.\ Worst Case}
Instead of studying the average success probability \eqref{eq:success probability} of a region estimator, we may also consider its  \emph{worst-case success probability}, which is defined as $p_\text{worst} = \inf_x \braket{\rho^x_B, M_B(\mathcal E_x)}$. Since $p_\text{succ} \geq p_\text{worst}$ with respect to any prior, we can maximize \eqref{eq:hypothesis-testing lower bound} over all priors $p_X$ to obtain a lower bound purely in terms of the given family of probe states $\{\rho^x_B\}$ and the desired worst-case success probability.
As this strategy is easily implemented for the lower bounds established in the following, we will focus on the Bayesian scenario throughout the remainder of this work.

On the other hand, we might also be interested in lower-bounding the \emph{average volume} reported by the estimator rather than the worst-case volume \eqref{eq:maximal volume}. For a fixed value $x$ of the unknown parameter, it is given by
\begin{equation*}
  V_\text{avg}(x) = \int_Y d\braket{\rho^x_B, M_B(y)} \mu_X(\mathcal E_y).
\end{equation*}
Since $V_\text{avg}(x) = \int_X d\mu_X(x') \braket{\rho^x_B, M_B(\mathcal E_{x'})}$ by the proof of \eqref{eq:max volume bound}, we obtain the following variant of \eqref{eq:hypothesis-testing lower bound} by proceeding as above:
\begin{equation}
  \label{eq:hypothesis-testing lower bound average x}
  \frac {V_\text{avg}(x)} {\abs X} \geq \beta_{p_{\text{succ}}}(\rho_{XB}, \frac {\id_X} {\abs X} \otimes \rho_B^x).
\end{equation}
Thus we may bound the ``worst-case average volume'' $\sup_x V_\text{avg}(x)$ simply by optimizing \eqref{eq:hypothesis-testing lower bound average x} over all $x$. On the other hand, if our goal is to bound the average volume with respect to the prior, which is given by
\begin{equation*}
  V_\text{avg}
  = \int d\mu_X(x) p^x_X V_\text{avg}(x),
  = \int_Y d\braket{\rho_B, M_B(y)} \mu_X(\mathcal E_y),
\end{equation*}
with $\rho_B = \int d\mu_X(x) p^x_X \rho^x_B$ the average probe state, then we may do so directly by using
\begin{equation}
  \label{eq:hypothesis-testing lower bound average}
  \frac {V_\text{avg}} {\abs X} \geq \beta_{p_{\text{succ}}}(\rho_{XB}, \frac {\id_X} {\abs X} \otimes \rho_B),
\end{equation}
In general,
$V_{\max} \geq \max_x V_\text{avg}(x) \geq V_\text{avg}$
(but see \cite[Theorem 1]{blumekohout_2012_robust}).

We now describe a different approach to bound average volumes, based on the intuition that the volume of the reported region is typically of the order of $V_\text{avg}$.
To make this precise, let $Y$ be the random variable describing the POVM measurement outcome of a given region estimator (for $x$ chosen according to the prior). Then the volume of the corresponding confidence region $V := \mu_X(\mathcal E_Y)$ is itself a random variable and satisfies
\begin{equation*}
  \PP(V \geq V_\text{avg}/\varepsilon)
  \leq
  \frac {\mathbb E[V]} {V_\text{avg}/\varepsilon} = \varepsilon
\end{equation*}
for any $\varepsilon > 0$ (by Chebyshev's inequality). By truncating those regions $\mathcal E_y$ that are too large, we may therefore construct a new region estimator with maximal volume $V_\text{avg}/\varepsilon$ for which the success probability is still at least $p_\text{succ} - \varepsilon$. So if we apply the hypothesis-testing lower bound for the maximal volume \eqref{eq:hypothesis-testing lower bound} to this new estimator, we obtain that
\begin{equation*}
  V_\text{avg} \geq \varepsilon \sup_{\sigma_B} \beta_{p_\text{succ} - \varepsilon}(\rho_{XB}, \id_X \otimes \sigma_B)
\end{equation*}
for any choice of $\varepsilon \in (0,p_\text{succ})$, e.g.\ for $\varepsilon = p_\text{succ} / 2$.
In this way, the results of this work, in particular our scaling results, imply directly corresponding lower bounds for the average volume, since they are all obtained by evaluating \eqref{eq:hypothesis-testing lower bound}.

\medskip\subsubsection*{Relationship to Information Theory}
As described in the introduction, the right-hand side of \eqref{eq:hypothesis-testing lower bound} is closely related to the conditional hypothesis-testing entropy as defined in the literature \cite{wangrenner12,hayashitomamichel12,generalizedentropies12}.
Moreover, \eqref{eq:hypothesis-testing lower bound} directly implies a lower bound in terms of the \emph{conditional von Neumann entropy}, defined as $H(X|B)_{\rho_{XB}} := -D(\rho_{XB} \Vert \id_X \otimes \rho_B)$ in terms of the relative entropy $D$ \cite{ohya_petz_1993_quantum},\footnote{For compact $X$, we can also define the conditional entropy as ${H(X|B) = -\int_X d\mu_X(x) D(\rho^x_B \Vert \rho_B)}$ \cite[Prop.~8]{berta_christandl_furrer_et_al_2013_continuous}.} that is closely related to Fano's inequality.

\begin{prp}
  \label{entropic lower bound}
  For any region estimator, we have the lower bound
  \begin{equation*}
    h(p_\text{succ}) + p_\text{succ} \log V_{\text{avg}} + (1 - p_\text{succ}) \log \abs X \geq H(X|B)_{\rho_{XB}},
  \end{equation*}
  where $h(p) = -p \log p-(1-p)\log (1-p)$ is the binary entropy function and $\rho_{XB}$ is the cq-state corresponding to the ensemble of probe states.
\end{prp}
\begin{IEEEproof}
  We start with the data-processing inequality for the relative entropy which, when applied to the optimal test for $\beta_\alpha = \beta_\alpha(\rho_{XB}, {\id_X} / {\abs X} \otimes \rho_B)$, shows that
  \begin{align*}
    -H(X|B)
    = D(\rho_{XB} \Vert \id_X \otimes \rho_B)
    \geq d \left( \alpha \Vert \beta_\alpha \right) - \log \abs X,
  \end{align*}
  where $d(p \Vert q) = p \log \frac p q + (1-p) \log \frac {1-p} {1-q}$ denotes the binary relative entropy function
  (see \eqref{eq:hypothesis test dpi} in Appendix~\ref{zero duality gap appendix}).
  Using the lower bound $d(p \Vert q) \geq -h(p) + p \log 1/q$ \cite[(156)]{polyanskiy_poor_verdu_2010_channel}, we find that
  \begin{equation*}
    -H(X|B) \geq -h(\alpha) + \alpha \log \frac 1 {\beta_\alpha} - \log \abs X.
  \end{equation*}
  For any region estimator, the hypothesis-testing lower bound \eqref{eq:hypothesis-testing lower bound average} now gives that
  \begin{equation*}
    -H(X|B) \geq -h(p_\text{succ}) + p_\text{succ} \log \frac {\abs X} {V_{\text{avg}}} - \log \abs X,
  \end{equation*}
  which is equivalent to the advertised lower bound.
\end{IEEEproof}

There are several ways in which the parameter estimation problem in the form considered here can be related to well-known tasks in quantum information theory.
For one, any region estimator can be understood as a list decoder \cite{sudan_2000_list} for joint source-channel coding, for the source described by the prior $p_X$ and the cq-channel $x \mapsto \rho^x_B$ induced by the family of probe states. Indeed, \eqref{eq:hypothesis-testing lower bound} can be understood as a quantum version of the list-decoding lower bound \cite[Theorem 4]{kostinaverdu13}, adapted to the case of a \emph{trivial} encoder. This observation was one of the starting points of this work.

Parameter estimation in general can also be related to source compression with side information. Here, $\rho^x_B$ is the side information, and the goal in parameter estimation is to decompress given \emph{only} the side information.
In fact, one can also establish the following, perhaps unexpected link to \emph{lossless} source compression. Suppose that the parameter space is finite (i.e., we are concerned with multiple hypothesis testing), and that we have a region estimator that reports $\delta$-balls with respect to some metric on $X$. Then we can turn $X$ into a graph by connecting any two points that have distance less than $\delta$ with respect to the metric. Such a graph can be colored with $V_{\max}$ colors, and the knowledge of the color can be used to pick out the correct point from any $\delta$-ball reported by the estimator with probability $p_{\text{succ}}$. This defines a lossless compression scheme for the source $p_X$ and side information $(p^x_B)$ which uses $m = \log_2 V_{\max}$ bits and which works with error probability at most $1-p_{\text{succ}}$. It is amusing to note that our hypothesis-testing lower bound \eqref{eq:hypothesis-testing lower bound} applied to this construction agrees precisely with the converse bound for source compression established in \cite[Theorem 9]{hayashitomamichel12}.

\medskip\subsubsection*{Mean-Square Error}
The bound \eqref{eq:hypothesis-testing lower bound} can also be used to derive statements about the mean-square error of point estimators.
Recall that a \emph{point estimator} is simply a POVM $M_B$ on the probe system $\mathcal H_B$ with outcomes in the parameter space $X$, which we assume is a metric measure space with metric $d_X$ (i.e., the measure $\mu_X$ is defined on the corresponding Borel $\sigma$-algebra).
Let $X$ and $\hat X$ denote random variables describing the prior and the estimate, respectively. That is, $X$ is distributed according to $p_X$, and $\hat X$ is the random variable with distribution $\braket{\rho^x_B, M_B(-)}$ when conditioned on $X = x$.
Then the \emph{mean square error} of the estimator is defined as
\begin{equation*}
  \Delta^2 := \mathbb E[d_X(X, \hat X)^2]. 
\end{equation*}

Now observe that we can also consider $M_B$ as a region estimator by interpreting its estimate as the center of a $\delta$-ball for some fixed radius $\delta > 0$. Mathematically, this corresponds to the choice $Y = X$ and $\mathcal E = \{ (x, \hat x) : d_X(x, \hat x) < \delta \}$.
Clearly, the average volume reported by this estimator can be upper-bounded by the maximal volume of a $\delta$-ball in $X$,
\begin{equation}
\label{eq:maximal volume point estimator}
  V_{\text{avg}} \leq V_{\max} \leq b_X(\delta) := \sup_{\hat x} \mu_X(\{ x : d_X(x, \hat x) < \delta \}).
\end{equation}
On the other hand, Chebyshev's inequality shows that
\begin{equation}
\label{eq:success probability point estimator}
  p_{\text{succ}} = 1 - \PP(d_X(X, \hat X) \geq \delta) \geq 1 - \frac {\Delta^2} {\delta^2}.
\end{equation}
Thus the hypothesis-testing lower bound \eqref{eq:hypothesis-testing lower bound} gives a constraint on the mean-square error
for any choice of $\delta > 0$. We will later apply this to recover the well-known Heisenberg and shot-noise scaling bounds for phase estimation (\autoref{subsec:phase-estimation}).
Conversely, any achievability result for the mean-square error can in this way be lifted to region estimation.
As a corollary of \autoref{entropic lower bound}, we can also prove the following entropic lower bound for the mean-square error:

\begin{cor}
  \label{mse entropic}
  For any point estimator with mean-square error $\Delta^2$, we have the lower bound
  \begin{equation*}
    b_X(\sqrt 2 \Delta) \geq \frac 1 {4 \abs X} e^{2 H(X|B)},
  \end{equation*}
  where $b_X(\delta)$ denotes the maximal volume of a $\delta$-ball as defined in \eqref{eq:maximal volume point estimator}.
\end{cor}
\begin{IEEEproof}
  Using \autoref{entropic lower bound}, \eqref{eq:success probability point estimator}, and \eqref{eq:maximal volume point estimator}, we get that
  \begin{align*}
    H(X|B)
    &\leq \log 2 + p_\text{succ} \log V_{\text{avg}} + (1 - p_\text{succ}) \log \abs X \\
    &\leq \log 2 + \left( 1 - \frac {\Delta^2} {\delta^2} \right) \log V_{\text{avg}} + \frac {\Delta^2} {\delta^2} \log \abs X \\
    &\leq \log 2 + \left( 1 - \frac {\Delta^2} {\delta^2} \right) \log b_X(\delta) + \frac {\Delta^2} {\delta^2} \log \abs X
  \end{align*}
  as long as we choose $\delta \geq \Delta$, since a convex combination of the two quantities $\log V_{\text{avg}}$ and $\log |X|$ can only increase if weight is shifted away from the lesser ($\log V_{\text{avg}}$). In particular, for $\delta = \sqrt 2 \Delta$ we find that
  \begin{equation*}
    H(X|B) \leq \log 2 + \frac 1 2 \log b_X(\delta) + \frac 1 2 \log \abs X,
  \end{equation*}
  which implies the asserted bound.
\end{IEEEproof}

For example, for phase estimation, where $X = U(1) = [0,2\pi]/\!\!\sim$ and $b_X(\delta) \leq 2 \delta$, we find that
\begin{equation*}
  \Delta \geq \frac 1 {16 \sqrt 2 \pi} e^{2 H(X|B)}.
\end{equation*}
For comparison, the well-known entropic lower bound for an arbitrary, not necessarily periodic, 
``shift parameter'' in $X = \RR$ that follows from rate-distortion theory \cite{nair12,hall_wiseman_2012} reads
\begin{equation*}
  \Delta \geq \frac 1 {2 \pi e} e^{2 H(X|\hat X)} \geq \frac 1 {2 \pi e} e^{2 H(X|B)}.
\end{equation*}

\medskip\subsubsection*{Properties of $\beta_\alpha$} In the remainder of this section we collect some useful properties of $\beta_\alpha$ that will later facilitate the computation of the hypothesis-testing lower bound.

Observe that the right-hand side of \eqref{cq hypothesis primal} is a linear cone program of locally convex topological vector spaces in duality (see, e.g., \cite[\S{}IV.6]{barvinok02}). The dual program is given by
\begin{equation}
\label{cq hypothesis dual}
\begin{aligned}
  \beta^*_\alpha(\rho_{XB}, \sigma_{XB}) = \sup \{ \alpha \mu - \braket{\tau_{XB}, \id_{XB}} : \\
  \mu \geq 0, \tau_{XB} \geq 0,
  \tau_{XB} \geq \mu \rho_{XB} - \sigma_{XB} \}.
\end{aligned}
\end{equation}
It can be established that the primal value, i.e.\ the infimum in \eqref{cq hypothesis primal} is always attained. Moreover, there is \emph{zero duality gap}: we have $\beta^*_\alpha = \beta_\alpha$ for all $\alpha \in [0, 1]$.
We note that the quantity $\beta_\alpha$ is monotonously increasing and continuous on $\alpha \in [0,1]$ (see Appendix~\ref{zero duality gap appendix}, where we establish these claims for hypothesis testing in general von Neumann algebras).

In the case where $X$ is a singleton space, \eqref{cq hypothesis primal} and \eqref{cq hypothesis dual} reduce to the usual primal and dual formulation for hypothesis testing between two quantum states $\rho_B$ and $\sigma_B$,
\begin{align}
\nonumber
   & \quad\beta_\alpha(\rho_B, \sigma_B) \\
\label{q hypothesis primal and dual}
   &= \inf \{ \braket{\sigma_B, E_B} : 0 \leq E_B \leq \id_B, \braket{\rho_B, E_B} \geq \alpha \} \\
\nonumber
   &= \sup \{ \alpha \mu - \braket{\tau_B, \id_B} : \mu \geq 0, \tau_B \geq 0, \tau_B \geq \mu \rho_B - \sigma_B \}.
\end{align}
It is easy to see that
\begin{equation}
\label{tests with same prior factorize}
  \beta_\alpha(p_X \otimes \rho_B, p_X \otimes \sigma_B) =
  \beta_\alpha(\rho_B, \sigma_B).
\end{equation}
for all probability densities $p_X$ and quantum states $\rho_B$, $\sigma_B$.

Moreover, we have the following \emph{data-processing inequality}: For any CPTP map $\Lambda$,
\begin{equation}
  \label{data processing}
  \beta_\alpha(\Lambda(\rho), \Lambda(\sigma)) \geq \beta_\alpha(\rho, \sigma),
\end{equation}
which follows easily from the fact that the dual channel $\Lambda^*$ is unital and positive, and therefore maps feasible tests onto feasible tests.


\begin{lem}
  \label{smoothing tests}
  Given any $\rho$, $\rho'$ for which $\tfrac 12 \norm{\rho-\rho'}_1 \leq \delta$ for some $\delta\geq 0$, then for any $\sigma$
  $$\beta_{\alpha+\delta}(\rho,\sigma)\geq \beta_\alpha(\rho',\sigma).$$
\end{lem}
\begin{IEEEproof}
  The claim follows by observing that any feasible test $E$ for $\beta_{\alpha+\delta}(\rho,\sigma)$ is also feasible for $\beta_{\alpha}(\rho',\sigma)$. Indeed, by the properties of the trace distance,
  \begin{align*}
    \delta &\geq \tfrac 12 \norm{\rho-\rho'}_1 \\
    &=\max_{0\leq P\leq \id} \braket{\rho-\rho', P}\\
    &\geq \braket{\rho, E}-\braket{\rho', E}\\
    &\geq (\alpha + \delta) - \braket{\rho', E},
  \end{align*}
  so $E$ is indeed feasible for $\beta_{\alpha}(\rho',\sigma)$.
\end{IEEEproof}

\begin{lem}
  \label{projective measurement lemma}
  Let $\rho$ be a quantum state, $(P_j)_{j=1}^J$ a projective measurement with $J$ outcomes, and $\sigma = \sum_j P_j \rho P_j$ the
  corresponding post-measurement state. Then, $\beta_\alpha(\rho, \sigma) \geq \alpha / J$.
\end{lem}
\begin{IEEEproof}
  Let $Z$ denote the operator that acts by multiplication with $\omega^j$ on the support of $P_j$, with $\omega$ a primitive $J$-th root of unity.
  Then we can write $\sigma$ as the group average
  \begin{equation*}
    \sigma = \frac 1 J \sum_{j=1}^J Z^j \rho (Z^j)^\dagger.
  \end{equation*}
  Clearly, $\frac 1 J \rho \leq \sigma$, so that $\mu = 1 / J$, $\tau = 0$ are feasible for the dual formulation of $\beta_\alpha(\rho, \sigma)$.
  We conclude that $\beta_\alpha(\rho, \sigma) \geq \alpha / J$ by (weak) duality.
\end{IEEEproof}

\section{Covariant estimation}
\label{sec:covariant}

In the following, let the parameter space $X$ be a smooth manifold equipped with the transitive action of a compact, connected Lie group $G$ and $G$-invariant metric $\mu_X$. Thus $\mu_X$ is induced by the unique Haar measure $\mu_G$ on $G$ with $\abs G = \abs X$. We moreover assume that the Hilbert space $\mathcal H_B$ is a unitary $G$-representation.
We will be interested in studying probe states $(\rho^x_B)$ that form a \emph{covariant family}, i.e.\ for which $\rho^{g x}_B = g \rho^x_B g^{-1}$.

Let $x_0 \in X$ be an arbitrary base point. Then we have that $X = G \cdot x_0 \cong G / K$, where $K$ is the stabilizer of $x_0$. It follows that the probe state $\rho^{x_0}_B$ has an additional symmetry,
\begin{equation}
\label{eq:stabilizer symmetry}
  k \rho^{x_0}_B k^{-1} = \rho^{x_0}_B
  \qquad
  (\forall k \in K)
\end{equation}
The stabilizer subgroup $K \subseteq G$ is uniquely determined up to conjugation.

\begin{prp}
  \label{covariant lower bound}
  For any region estimator for a covariant family $(\rho^x_B)$ and prior $p_X$, we have the lower bound
  \begin{equation}
  \label{eq:covariant lower bound}
    \frac {V_{\max}} {\abs X} \geq \sup_{\tilde\sigma_B} \beta_{p_{\text{succ}}}(p_X \otimes \rho^{x_0}_B, \frac {\id_X} {\abs X} \otimes \tilde\sigma_B),
  \end{equation}
  where the supremum runs over all $G$-invariant states $\tilde\sigma_B$; $x_0$ is an arbitrary base point.
\end{prp}
\begin{IEEEproof}
  To see this, let $E_{XB} = (E^x_B)$ be a feasible test for $\beta_{p_{\text{succ}}}(\rho_{XB}, \id_X / \abs X \otimes \tilde\sigma_B)$.
  We define the ``untwisted test'' $\tilde E_{XB} = (\tilde E^x_B)$ by $\tilde E^x_B := g^{-1} E^x_B g$ where $g = g(x)$ is chosen such that $g x_0 = x$ \footnote{More conceptually, $\tilde E_{XB}$ could also be defined by integrating $g^{-1} E^{g x_0}_B g$ over the fibers of $G \rightarrow X, g \mapsto g x_0$.}.
  Then,
  \begin{align*}
    &\quad\braket{p_X \otimes \rho^{x_0}_B, \tilde E_{XB}}
    = \int d\mu_X(x) \, p^{x}_X \braket{\rho^{x_0}_B, \tilde E^x_B} \\
    &= \int d\mu_X(x) \, p^{x}_X \braket{\rho^x_B, E^x_B}
    = \braket{\rho_{XB}, E_{XB}}.
  \end{align*}
  On the other hand, by using the $G$-invariance of $\tilde\sigma_B$ we find
  \begin{align*}
      &\braket{\id_X \otimes \tilde\sigma_B, \tilde E_{XB}}
    = \int d\mu_X(x) \braket{\tilde\sigma_B, \tilde E^x_B} \\
    = &\int d\mu_X(x) \braket{\tilde\sigma_B, E^x_B}
    = \braket{\id_X \otimes \tilde\sigma_B, E_{XB}}.
  \end{align*}
  Therefore,
  \begin{equation*}
    \beta_{p_{\text{succ}}}(\rho_{XB}, \frac {\id_X} {\abs X} \otimes \tilde\sigma_B)
     \geq
    \beta_{p_{\text{succ}}}(p_X \otimes \rho^{x_0}_B, \frac {\id_X} {\abs X} \otimes \tilde\sigma_B),
  \end{equation*}
  and hence the claim follows from \eqref{eq:hypothesis-testing lower bound}.
\end{IEEEproof}

\begin{cor}
	\label{cor:uniformprior-statedep}
  For any region estimator for a covariant family $(\rho^x_B)$ and invariant (hence uniform) prior, we have
  \begin{align}
  \label{eq:invariant lower bound supinf}
    &\frac {V_{\max}} {\abs X}
    \geq \sup_{\tilde\sigma_B} \beta_{p_\text{succ}}(\rho^{x_0}_B, \tilde\sigma_B) \\
  \label{eq:invariant lower bound infsup}
    = &\inf \{ \norm{E_B^G}_\infty : E_B \in \mathcal O_B, 0 \leq E_B \leq \id_B, \\
  \nonumber
  &\qquad\qquad\qquad\braket{\rho_B^{x_0}, E_B} \geq p_{\text{succ}} \},
  \end{align}
  where the supremum in \eqref{eq:invariant lower bound supinf} runs over all $G$-invariant states $\tilde\sigma_B$, $x_0$ is an arbitrary point in $X$, and $E_B^G$ denotes the $G$-average of the operator $E_B$.
\end{cor}
\begin{IEEEproof}
  Since the action is transitive, any invariant prior is equal to the uniform prior, so that $p_X = \id_X / \abs X$.
  Thus the first lower bound follows from combining \eqref{tests with same prior factorize} and \eqref{eq:covariant lower bound}.

  We now compute the supremum: By Fan's minimax theorem \cite[Theorem 2]{fan53},
  \begin{align*}
      &\sup_{\tilde\sigma_B} \beta_{p_\text{succ}}(\rho^{x_0}_B, \tilde\sigma_B) \\
    = &\sup_{\tilde\sigma_B} \min \{ \braket{\tilde\sigma_B, E_B} : 0 \leq E_B \leq \id_B, \braket{\rho^{x_0}_B, E_B} \geq {p_\text{succ}} \} \\
    = &\min \{ \sup_{\tilde\sigma_B} \braket{\tilde\sigma_B, E_B} : 0 \leq E_B \leq \id_B, \braket{\rho^{x_0}_B, E_B} \geq {p_\text{succ}} \}
  \end{align*}
  since the set of feasible tests is weak-$\star$-compact. But
  \begin{equation*}
    \sup_{\tilde\sigma_B} \braket{\tilde\sigma_B, E_B} =
    \sup_{\tilde\sigma_B} \braket{\tilde\sigma_B, E_B^G} =
    \sup_{\sigma_B} \braket{\sigma_B, E_B^G} =
    \norm{E_B^G}_\infty,
  \end{equation*}
  where the last supremum is taken over all states $\sigma_B$.
\end{IEEEproof}

\begin{exl}
  In general, the optimal $\tilde\sigma_B$ in \eqref{eq:invariant lower bound supinf} is not given by the $G$-average of the probe state.
  Consider e.g.\ the action of $U(1) \subseteq \SU(2)$ on $\CC^2$ generated by the Pauli $\sigma_z$-operator, with $\rho_B^{x_0}$ the pure state $\sqrt{0.9} \ket 0 + \sqrt{0.1} \ket 1$.
  For $p_\text{succ} = 1$, the optimal test $E_B$ is given precisely by $\rho_B^{x_0}$, so that $E_B^G = (\rho_B^{x_0})^G = 0.9 \proj 0 + 0.1 \proj 1$ and $\beta_1(\rho_B^{x_0}, \tilde\sigma_B) = 0.9 \braket{0 | \tilde\sigma_B | 0} + 0.1 \braket{0 | \tilde\sigma_B | 1}$
  Clearly, the optimal $\tilde\sigma_B$ is not the $G$-average of $\rho_B^{x_0}$ but rather the pure state $\proj 0$.
  By continuity the same conclusion holds for $p_{\text{succ}} \approx 1$.
\end{exl}

We now give a \emph{state-independent} lower bound for the volume of any confidence region for a given confidence level. Such a lower bound is an ultimate limit to the precision of any region estimator for an arbitrary covariant ensemble on $\mathcal H_B$.
From now on we will assume that $\mathcal H_B$ is finite-dimensional.

Before we state the result, let us consider the \emph{isotypical decomposition} of $\mathcal H_B$ with respect to the group $G$, i.e.\ the decomposition
\begin{equation}
\label{eq:isotypical G}
  \mathcal H_B \cong \bigoplus_\lambda V_{G,\lambda} \otimes \CC^{m_\lambda},
\end{equation}
where $m_\lambda$ denotes the multiplicity of an irreducible representation $V_{G,\lambda}$ of $G$.
Fix a base point $x_0 \in X$ and denote by $K \subseteq G$ the stabilizer of $x_0$.
By decomposing each irreducible $G$-representation $V_{G,\lambda}$ into irreducible $K$-representations $V_{K,\mu}$, we obtain from \eqref{eq:isotypical G} that
\begin{equation}
\label{eq:isotypical K}
\begin{aligned}
  \mathcal H_B
  \cong &\bigoplus_\lambda \left( \bigoplus_\mu V_{K,\mu} \otimes \CC^{m^\lambda_\mu} \right) \otimes \CC^{m_\lambda} \\
  = &\bigoplus_\mu V_{K,\mu} \otimes \left( \bigoplus_\lambda \CC^{m^\lambda_\mu} \otimes \CC^{m_\lambda} \right),
\end{aligned}
\end{equation}
where $m^\lambda_\mu$ denotes the multiplicity of $V_{K,\mu}$ in $V_{G,\lambda}$.

\begin{cor}
  \label{state-independent invariant lower bound}
  For any region estimator for a covariant family and uniform prior, we have
  \begin{equation}
  \label{eq:state-independent invariant lower bound}
    \frac {V_{\max}} {\abs X} \geq
    \inf_{\rho^{x_0}_B} \sup_{\tilde\sigma_B} \beta_{p_\text{succ}}(\rho^{x_0}_B, \tilde\sigma_B) =
    \frac {p_\text{succ}} {\max_\mu \sum_\lambda d_\lambda r^\lambda_\mu / d_\mu},
  \end{equation}
  where the infimum runs over all $K$-invariant states $\rho^{x_0}_B$ and the supremum over all $G$-invariant states $\tilde\sigma_B$ on $\mathcal H_B$.
  Moreover, $d_\lambda$ and $d_\mu$ denote the dimension of $V_{G,\lambda}$ and $V_{K,\mu}$, respectively, and $r^\lambda_\mu := \min \{ m^\lambda_\mu, m_\lambda \}$, where we use the same notation as in the decompositions \eqref{eq:isotypical G} and \eqref{eq:isotypical K}.

  In particular, if $X = G$ is a group then we have the lower bound
  \begin{equation}
  \label{eq:state-independent invariant lower bound for groups}
    \frac {V_{\max}} {\abs X} \geq
    \frac {p_\text{succ}} {\sum_\lambda d_\lambda r_\lambda},
  \end{equation}
  where $r_\lambda := \min \{ d_\lambda, m_\lambda \}$.
\end{cor}
\begin{IEEEproof}
  Recall from \eqref{eq:stabilizer symmetry} that the probe state $\rho^{x_0}_B$ is necessarily $K$-invariant.
  Thus we get a state-independent lower bound by optimizing \eqref{eq:invariant lower bound infsup} over all
  $K$-invariant probe states $\rho^{x_0}_B$.
  \begin{align*}
    &\frac {V_{\max}} {\abs X}
    \geq \inf_{\rho^{x_0}_B} \sup_{\tilde\sigma_B} \beta_{p_\text{succ}}(\rho^{x_0}_B, \tilde\sigma_B) \\
    = &\inf_{\rho^{x_0}_B} \inf \{ \norm{E_B^G}_\infty : 0 \leq E_B \leq \id_B, \braket{\rho^{x_0}_B, E_B} \geq p_\text{succ} \} \\
    = &\inf \{ \norm{E_B^G}_\infty : 0 \leq E_B \leq \id_B, \max_{\rho^{x_0}_B} \braket{\rho^{x_0}_B, E_B} \geq p_\text{succ} \} \\
   = &\inf \{ \norm{E_B^G}_\infty : 0 \leq E_B \leq \id_B, \norm{E_B^K}_\infty \geq p_\text{succ} \} \\
   = &p_\text{succ} \inf \{ \norm{E_B^G}_\infty : 0 \leq E_B = E_B^K, \norm{E_B}_\infty = 1 \}.
  \end{align*}
  By Schur's lemma, each such $E_B$ can be written in the form $E_B = \bigoplus_\mu \id_{V_{K,\mu}} \otimes E_\mu$ with respect to the isotypical decomposition \eqref{eq:isotypical K}. We may in fact assume that
  \begin{equation*}
    E_B
    = \id_{V_{K,\mu}} \otimes \proj{\psi_\mu}
    = d_\mu \frac {\id_{V_{K,\mu}}} {d_\mu} \otimes \proj{\psi_\mu}.
  \end{equation*}
  Indeed, restricting to a single summand and replacing $E_\mu$ by the rank-one projector onto a maximal eigenvector will never increase $\norm{E_\mu^G}_\infty$.
  Now decompose $\ket{\psi_\mu} = \sum_\lambda \sqrt{p_\lambda} \ket{\psi_\lambda}$ according to the direct sum $\bigoplus_\lambda \CC^{m^\lambda_\mu} \otimes \CC^{m_\lambda}$, and denote by $\rho_\lambda$ the reduced density matrix of $\ket{\psi_\lambda}$ on $\CC^{m_\lambda}$.
  It follows from another application of Schur's lemma that
  \begin{equation*}
    E_B^G = d_\mu \sum_\lambda p_\lambda \frac {\id_{V_{G,\lambda}}} {d_\lambda} \otimes \rho_\lambda.
  \end{equation*}
  By the Schmidt decomposition, the rank of $\rho_\lambda$ is at most $r^\lambda_\mu$, while its trace is one. Therefore,
  \begin{align*}
    &\norm{E_B^G}_\infty
    = d_\mu \max_\lambda p_\lambda \frac 1 {d_\lambda} \norm{\rho_\lambda}_\infty \\
    &\geq \max_\lambda p_\lambda \frac {d_\mu} {d_\lambda} \frac 1 {r^\lambda_\mu}
    \geq \frac 1 {\sum_\lambda d_\lambda r^\lambda_\mu / d_\mu}
  \end{align*}
  By minimizing over $\mu$ we arrive at the advertised lower bound.
\end{IEEEproof}

By examining the final inequalities in the above proof, it is easy to extract the form of probe states $\rho^{x_0}_B$ for which the state-dependent lower bound \eqref{eq:invariant lower bound supinf} attains the universal lower bound \eqref{eq:state-independent invariant lower bound}. E.g., in the case where $X=G$ we can choose $\rho_B = \proj\psi_B$ with $\ket\psi_B = \sum_\lambda z_\lambda \ket{\psi_\lambda}_B$, $\abs{z_\lambda}^2 = \frac {d_\lambda r_\lambda} {\sum_{\lambda'} d_{\lambda'} r_{\lambda'}}$ and $\ket{\psi_\lambda}_B = \frac 1 {\sqrt {r_\lambda}} \sum_{m=1}^{r_\lambda} \ket{m}_{V_\lambda} \otimes \ket{m}_{\CC^{m_\lambda}}$ (cf.\ \cite{gourmarvianspekkens09}, where it was shown that such states also achieve ``maximal $G$-asymmetry'' as defined in that work, and \cite{chiribella_dariano_sacchi_2005_optimal}, where it was shown that states of the general form $\sum_\lambda z_\lambda \ket{\psi_\lambda}_B$ are optimal for group element estimation with respect to a wide class of risk functions).

Although we have so far established the lower bounds \eqref{eq:invariant lower bound supinf} and \eqref{eq:state-independent invariant lower bound} for uniform priors only, it is easy to generalize these to general priors by adapting a chain rule proved in \cite{generalizedentropies12}:

\begin{lem}
  \label{prior decomposition lemma}
  Let $p_X \in L^1(X, \mu_X)$ be a probability density, $\rho_B$ and $\sigma_B$ be quantum states on $\mathcal H_B$, and $\alpha \geq 0$, $\alpha' > 0$ such that $\alpha + \sqrt{2\alpha'} \leq 1$. Then we have
  \begin{align*}
    \beta_{\alpha+\sqrt{2\alpha'}}(p_X \otimes \rho_B, \frac {\id_X} {\abs X} \otimes \sigma_B)
    \geq \beta_\alpha(p_X, \frac {\id_X} {\abs X})
    \cdot
    \frac 1 {\alpha'} \beta_{\alpha'}(\rho_B, \sigma_B).
  \end{align*}
\end{lem}
\begin{IEEEproof}
  We closely follow the proof of \cite[Proposition 5.1]{generalizedentropies12}.
  Let $\mu_X$, $\tau_X$ be feasible for the dual formulation of $\beta_\alpha(p_X, \id_X / \abs X)$,
  and let $\mu_B$, $\tau_B$ be feasible for the dual formulation of $\beta_{\alpha'}(\rho_B, \sigma_B)$. Then,
  \begin{align*}
    &\quad \mu_B \mu_X p_X \otimes \rho_B \\
    &\leq \mu_B \left( \frac {\id_X} {\abs X} + \tau_X \right) \otimes \rho_B \\
    &= \frac {\id_X} {\abs X} \otimes \left( \mu_B \rho_B \right) + \mu_B \tau_X \otimes \rho_B \\
    &\leq \frac {\id_X} {\abs X} \otimes \left( \sigma_B + \tau_B \right) + \mu_B \tau_X \otimes \rho_B
  \end{align*}
  Define $T_B := \sigma_B^{1/2} \left( \sigma_B + \tau_B \right)^{-1/2}$, where the inverse is taken on $\supp \sigma_B \subseteq \supp \sigma_B + \tau_B$ \footnote{That is, $\left( \sigma_B + \tau_B \right)^{-1}$ is by definition the inverse of the positive operator $P \left( \sigma_B + \tau_B \right) P$ on its support, where $P$ denotes the orthogonal projection onto $\supp \sigma_B$.}. Conjugating the above operator inequality with $T_B$, we find that
  \begin{align*}
    \mu_B \mu_X p_X \otimes T_B \rho_B T_B^\dagger \leq \frac {\id_X} {\abs X} \otimes \sigma_B + \mu_B \tau_X \otimes T_B \rho_B T_B^\dagger.
  \end{align*}
  Thus $\mu = \mu_B \mu_X$, $\tau_{XB} = \mu_B \tau_X \otimes T_B \rho_B T_B^\dagger$ are feasible for the dual formulation of the
  hypothesis test between $p_X \otimes T_B \rho_B T_B^\dagger$ and $\id_X / \abs X \otimes \sigma_B$, so that
  \begin{align*}
    &\quad \beta_\alpha(p_X \otimes T_B \rho_B T_B^\dagger, \frac {\id_X} {\abs X} \otimes \sigma_B) \\
    &\geq \mu_B \mu_X \alpha - \braket{\mu_B \tau_X \otimes T_B \rho_B T_B^\dagger, \id_{XB}} \\
    &= \mu_B \left( \mu_X \alpha - \braket{\tau_X \otimes \rho_B T_B^\dagger T_B, \id_{XB}} \right) \\
    &\geq \mu_B \left( \mu_X \alpha - \braket{\tau_X, \id_X} \right) \\
    &\geq \frac 1 {\alpha'} \left( \mu_B \alpha' - \braket{\tau_B, \id_B} \right) \left( \mu_X \alpha - \braket{\tau_X, \id_X} \right) \\
  \end{align*}
  for any $\alpha' \in [0, 1]$. Here we have used that $T_B$ is a contraction, i.e.\ that $T_B^\dagger T_B \leq \id_B$.
  By optimizing over all dual feasible points, we find that
  \begin{equation}
    \label{eq:presmoothing}
    \beta_\alpha(p_X \otimes T_B \rho_B T_B^\dagger, \frac {\id_X} {\abs X} \otimes \sigma_B)
    \geq \beta_\alpha(p_X, \frac {\id_X} {\abs X}) \frac 1 {\alpha'} \beta_{\alpha'}(\rho_B, \sigma_B).
  \end{equation}
  On the other hand, we have that
  \begin{equation*}
    \frac 1 2 \norm{p_X \otimes T_B \rho_B T_B^\dagger - p_X \otimes \rho_B}_1
    = \frac 1 2 \norm{\rho_B - T_B \rho_B T_B^\dagger}_1 \leq \sqrt{2 \alpha'}.
  \end{equation*}
  where the inequality is established just like in the proof of \cite[Proposition 5.1]{generalizedentropies12}.
  Thus the claim follows from \eqref{eq:presmoothing} and \autoref{smoothing tests}.
\end{IEEEproof}

\begin{thm}
  \label{state-independent covariant lower bound}
  For any region estimator for a covariant family and prior $p_X$, we have the lower bound
  \begin{equation}
  \label{eq:state-independent covariant lower bound}
    \frac {V_{\max}} {\abs X} \geq \beta_{p_\text{succ}}(p_X, \frac {\id_X} {\abs X}) \, \frac 1 {\max_\mu \sum_\lambda d_\lambda r^\lambda_\mu / d_\mu}.
  \end{equation}
  In particular, if $X = G$ is a group then we have the lower bound
  \begin{equation}
  \label{eq:state-independent covariant lower bound for groups}
    \frac {V_{\max}} {\abs X} \geq
    \beta_{p_\text{succ}}(p_X, \frac {\id_X} {\abs X}) \, \frac {1} {\sum_\lambda d_\lambda r_\lambda}.
  \end{equation}
  Here, $d_\lambda$, $d_\mu$, $r^\lambda_\mu$ and $r_\lambda$ are defined as in the statement of \autoref{state-independent invariant lower bound}.
\end{thm}
\begin{IEEEproof}
  For all $\alpha' > 0$ small that
  \begin{align*}
      \frac {V_{\max}} {\abs X}
    &\geq \sup_{\tilde\sigma_B} \beta_{p_\text{succ}}(p_X \otimes \rho_B^{x_0}, \frac {\id_X} {\abs X} \otimes \tilde\sigma_B) \\
    &\geq \beta_{p_\text{succ} - \sqrt{2\alpha'}}(p_X, \frac {\id_X} {\abs X}) \cdot \sup_{\tilde\sigma_B} \frac 1 {\alpha'} \beta_{\alpha'}(\rho_B^{x_0}, \tilde\sigma_B) \\
    &\geq \beta_{p_\text{succ} - \sqrt{2\alpha'}}(p_X, \frac {\id_X} {\abs X}) \frac 1 {\sum_\lambda d_\lambda r_\lambda},
  \end{align*}
  where we have used \eqref{eq:covariant lower bound}, \autoref{prior decomposition lemma}, and the identity in \eqref{eq:state-independent invariant lower bound} (in this order). Now let $\alpha' \rightarrow 0$ and use continuity of $\beta_\alpha$.
\end{IEEEproof}

Observe that \eqref{eq:state-independent covariant lower bound} and \eqref{eq:state-independent covariant lower bound for groups} reduce to \eqref{eq:state-independent invariant lower bound} and \eqref{eq:state-independent invariant lower bound for groups} in the case of a uniform prior.

\section{Asymptotics}
\label{sec:asymptotics}

We will now analyze the scaling of the lower bound \eqref{eq:state-independent covariant lower bound} when the probe system is a tensor power $\mathcal H_{B^N} = \mathcal H_B^{\otimes N}$ of a fixed, finite-dimensional representation $\mathcal H_B$.
Physically, this corresponds to the case where the probe states are generated by symmetric single-body operators, e.g.\ single-body Hamiltonians in the important case of $U(1)$-phase estimation.
We shall only treat the case where $X = G$ (but see \autoref{subsec:state estimation}).

\begin{lem}
  \label{heisenberg counting lemma}
  Let $T \subseteq G$ be a maximal torus (i.e.\ a maximal compact, connected, abelian subgroup). Then:
  \begin{enumerate}
    \item The number of isotypical components in $\mathcal H_B^{\otimes N}$ is $O(N^{\dim T})$.
    \item Each irreducible representation that occurs in $\mathcal H_B^{\otimes N}$ has dimension $O(N^{(\dim G - \dim T)/2})$.
  \end{enumerate}
\end{lem}
\begin{IEEEproof}[Proof for $G = T = U(1)$]
  For the first claim, we need to show that the Hamiltonian generating the $U(1)$-action on $\mathcal H_B$ has at most linearly many eigenvalues. After a choice of basis we may assume that $\mathcal H_B = \CC^d$, and that the $U(1) = [0,2\pi]/\!\!\sim$-action is generated by a diagonal Hamiltonian $H = \diag \vec h$ with integral entries $\vec h \in \ZZ^d$. Thus the action on $\mathcal H_B^{\otimes N}$ is generated by the one-body Hamiltonian $H_N = H \otimes \id + \ldots + \id \otimes H$. Clearly, $H_N$ is diagonal in the computational product basis, and the eigenvalue of a basis vector $\ket{\vec x}$ is equal to the inner product $\braket{\vec\omega, \vec h}$, where $\vec\omega \in \ZZ^d$ is the \emph{type} of $\vec x$, specifying the number of occurrences of the symbols $1, \ldots, d$ in a string $\vec x$. But $\braket{\vec\omega, \vec h}$ is an integer such that
  \begin{equation*}
    \abs{\braket{\vec\omega, \vec h}} \leq
    \norm{\vec\omega}_1 \norm{\vec h}_\infty =
    N \norm{\vec h}_\infty.
  \end{equation*}
  It follows that there are at most $2 N \norm{\vec h}_\infty + 1$ eigenvalues.

  For the second claim, recall that the irreducible representations of abelian groups are one-dimensional.
\end{IEEEproof}
\begin{IEEEproof}[Proof for general $G$]
  We will use some basic notions of the theory of compact Lie groups \cite{fultonharris91,cartersegalmacdonald95,kirillov08}.
  Without loss of generality we may assume that $G$ is semisimple, since we can always treat the connected part of the center via the above proof for $U(1)$.
  Let us denote by $\mathfrak t^*$ the dual of the Lie algebra $\mathfrak t$ of $T$, equipped with the inner product $(-,-)$ induced by the Killing form.
  We can choose a finite set of positive roots $R_+ = \{\alpha\} \subseteq \mathfrak t^*$.
  They span a proper cone; the dual cone with respect to the Killing form is called the positive Weyl chamber and denoted by $\mathfrak t^*_+$.
  The Weyl vector $\rho = \frac 1 2 \sum_\alpha \alpha$ is an element in the interior of both cones.
  We can use it to define a partial order on $\mathfrak t^*_+$: $\xi \succeq \xi'$ if and only if $\braket{\xi,\rho} \geq \braket{\xi',\rho}$. In particular, $\xi \succeq 0$ for all $\xi \in \mathfrak t^*_+$.
  There is a lattice $\Lambda^* \subseteq \mathfrak t^*$, called the weight lattice, which corresponds to the generators of one-parameter subgroups $U(1) \subseteq T$. The intersection $\Lambda^*_+ = \Lambda^* \cap \mathfrak t^*_+$ is called the set of dominant weights.
  The fundamental theorem of the representation theory of compact, connected Lie groups states that the irreducible representations of $G$ are labeled by an element $\lambda \in \Lambda^*_+$, called the \emph{highest weight}. In the familiar case where $G = \SU(2)$, $\Lambda^*_+$ can be identified with the set of non-negative half-integers, and $\lambda$ is the spin $j$ of the representation.

  Let us now consider the tensor product of two irreducible representations, and decompose it into irreducible representations,
  \begin{equation*}
    V_{G,\lambda} \otimes V_{G,\mu} = \bigoplus_\nu V_{G,\nu} \otimes \CC^{c^\nu_{\lambda,\mu}}.
  \end{equation*}
  Then it is well-known that $\lambda + \mu$ is the highest weight in this decomposition. That is, $\lambda + \mu \succeq \nu$ for all $\nu$ with respect to the order defined above. This generalizes the fact that the sum of the two spins is the largest term in the Clebsch--Gordan series for $\SU(2)$.

  Now let $\mathcal H_B = \bigoplus_{j=1}^J V_{G,\lambda_j} \otimes \CC^{m_j}$ be the isotypical decomposition of $\mathcal H_B$. Then,
  \begin{equation*}
    \mathcal H_B^{\otimes N} =
    \bigoplus_{j_1,\ldots,j_N} V_{G,\lambda_{j_1}} \otimes \ldots \otimes V_{G,\lambda_{j_N}} \otimes \CC^{m_{j_1} \cdots m_{j_N}}.
  \end{equation*}
  Set $\tilde\lambda := \sum_j \lambda_j$. Then $\tilde\lambda \succeq \lambda_j$ for all $j$, and it follows that $N \tilde\lambda \succeq \nu$ for all irreducible representations $V_{G,\nu}$ that appear in $\mathcal H_B^{\otimes N}$. Indeed, if we consider a tensor product of $N$ irreducible representations then the highest weight is given by the sum of the highest weights of the $N$ factors, which is always less than $N \tilde\lambda$. Therefore, the number of distinct irreducible representations that occur in $\mathcal H_B^{\otimes N}$ be upper-bounded by the cardinality of the set
  \begin{equation*}
    \Lambda^*_N := \{ \nu \in \Lambda^*_+ : \nu \preceq N \tilde\lambda \},
  \end{equation*}
  which scales at most as $N^{\dim T}$, since $\dim T = \dim \mathfrak t^*$ is the dimension of the weight lattice.
  This establishes the first claim.

  For the second claim, recall that the \emph{Weyl dimension formula} asserts that the dimension of an irreducible representation $V_{G,\lambda}$ is given by the polynomial
  \begin{equation}
    \label{eq:weyl dimension formula}
    p(\lambda) = \prod_{\alpha \in R_+} {(\alpha, \lambda + \rho)} / {(\alpha, \rho)}.
  \end{equation}
  The degree of $p(\lambda)$ is equal to $\abs{R_+}$, the number of positive roots, so that
  \begin{equation*}
    p(\nu) \leq N^{\abs{R_+}} \underbrace{\sup \{ p(\nu) : \nu \in \mathfrak t^*_+, \nu \preceq \tilde\lambda \}}_{< \infty}
  \end{equation*}
  for all $\nu \in \Lambda^*_N$. The second claim follows from this, since $\abs{R_+} = (\dim G - \dim T)/2$.
\end{IEEEproof}

\begin{thm}[Heisenberg limit]
\label{state-independent covariant heisenberg bound}
  For any region estimator for a covariant family on $\mathcal H_B^{\otimes N}$ and prior $p_G$, we have that
  \begin{equation}
  \label{eq:state-independent covariant heisenberg bound}
    \frac {V_{\max}} {\abs G} \geq C \, \beta_{p_\text{succ}}(p_G, \frac {\id_G} {\abs G}) \, N^{-\dim G}
  \end{equation}
  for $N$ large, where the constant $C > 0$ only depend on the representation $\mathcal H_B$.
\end{thm}
\begin{IEEEproof}
  By \autoref{heisenberg counting lemma}, we can estimate the right-hand side quantity in \eqref{eq:state-independent covariant lower bound for groups} by
  \begin{equation*}
    \sum_\lambda d_\lambda r_\lambda \leq \sum_\lambda d_\lambda^2 \leq O(N^{\dim T} N^{\dim G - \dim T}) = O(N^{\dim G}).
  \end{equation*}
  Thus the assertion is a consequence of the bound \eqref{eq:state-independent covariant lower bound for groups}.
\end{IEEEproof}

\autoref{state-independent covariant heisenberg bound} is the analog of Heisenberg scaling for region estimators. It provides an ultimate lower bound for any region estimator and family of probe states that is covariant for the tensor product action on $\mathcal H_B^{\otimes N}$.
In fact, by interpreting point estimators as region estimators and using \eqref{eq:maximal volume point estimator} and \eqref{eq:success probability point estimator}, we obtain Heisenberg scaling for the mean-square error as a direct consequence of our bound, generalizing results in the literature for $U(1)$ \cite{giovannettilloydmaccone06} and $\SU(d)$ \cite{chiribella_dariano_sacchi_2005_optimal,kahn_2007_fast} (see \autoref{sec:example} for worked examples).

\medskip\subsubsection*{Separable States}
We will now show that for a covariant family of separable probe states and abelian $G$, any region estimator satisfies a stronger lower bound than the one just established:

\begin{prp}
	\label{prp:separable}
  Let $G = T$ be abelian. Let $\rho_{B^N}$ be a (fully) separable state on $\mathcal H_B^{\otimes N}$.
  Then there exists a constant $D > 0$, only depending on the representation $\mathcal H_B$, such that
  \begin{equation*}
    \sup_{\tilde\sigma_{B^N}} \beta_\alpha(\rho_{B^N}, \tilde\sigma_{B^N}) \geq D {\alpha^{\dim T + 1}} {N^{-\dim T/2}}.
  \end{equation*}
  for all $\alpha > 0$, where the supremum runs over all $T$-invariant states $\tilde\sigma_{B^N}$.
\end{prp}
\begin{IEEEproof}
  We first consider the case where $\rho_{B^N}$ is a pure product state $\ket{\psi_{B^N}} = \ket{\psi_{B_1}} \otimes \ldots \otimes \ket{\psi_{B_N}}$.
  Recall that any compact abelian Lie group is a torus $T = U(1)^k$, where $k = \dim T$.
  Let us choose generators $H_1, \ldots, H_k$ of the action of each $U(1)$-factor of $\mathcal H_B$.
  The generators commute and can therefore be jointly diagonalized, i.e.\ there exists a decomposition $\mathcal H_B = \bigoplus_{\vec\omega} \mathcal H_B^{\vec\omega}$ into joint eigenspaces, with $\vec\omega \in \ZZ^k$ encoding the (integral) eigenvalues. The vector $\vec\omega$ is commonly called a \emph{weight}. Let us decompose each tensor factor of $\ket{\psi_{B^N}}$ accordingly, $\ket{\psi_{B_n}} = \sum_{\vec\omega} \sqrt{p_{n,\vec\omega}} \ket{n,\vec\omega}$, where each $\ket{n,\vec\omega} \in \mathcal H_B^{\vec\omega}$. Thus,
  \begin{align*}
    \ket{\psi_{B^N}} &= \sum_{\vec\omega_1,\ldots,\vec\omega_N} \sqrt{p_{1,\vec\omega_1} \ldots p_{N,\vec\omega_N}} \ket{1,\vec\omega_1} \otimes \ldots \otimes \ket{N,\vec\omega_N} \\
    &= \sum_{\vec\omega_1,\ldots,\vec\omega_N} \sqrt{p_{\vec\omega_1, \ldots, \vec\omega_N}} \ket{\vec\omega_1, \ldots, \vec\omega_N}
  \end{align*}
  where $p_{\vec\omega_1, \ldots, \vec\omega_N} := p_{1,\vec\omega_1} \cdots p_{N,\vec\omega_N}$ can be considered as the probability distribution of $N$ independent random variables $X_1$, \ldots, $X_N$, each with values in the finite set $\{\vec\omega\}$ of possible weights.

  Note that each $\ket{\vec\omega_1, \ldots, \vec\omega_N}$ is a joint eigenvector of the $T$-action on $\mathcal H_B^{\otimes N}$, which is generated by the single-body Hamiltonians $H_{N,j} = H_j \otimes \id + \ldots + \id \otimes H_j$. Clearly, the weight of $\ket{\vec\omega_1, \ldots, \vec\omega_N}$ is given by the sum of the individual weights, $\sum_n \vec\omega_n$.
  Let us thus consider the set of all eigenvectors whose weight is less than $\varepsilon$ away from the mean $\vec m := \mathbb E[X_1 + \ldots + X_n] = \sum_n \sum_{\vec\omega} p_{n,\vec\omega} \, \vec\omega$,
  \begin{equation*}
    \Omega_\varepsilon = \{ (\vec\omega_1, \ldots, \vec\omega_N) : \norm{\sum_{n=1}^N \vec\omega_n - \vec m}_2 < \varepsilon \}.
  \end{equation*}
  The constant $\varepsilon > 0$ will later be chosen appropriately.
  Then we can lower-bound the overlap between $\ket{\psi_{B^N}}$ and its normalized truncation
  $\ket{\psi'_{B^N}} \propto \sum_{(\vec\omega_j) \in \Omega_\varepsilon} \sqrt{p_{\vec\omega_1,\ldots,\vec\omega_N}} \ket{\vec\omega_1, \ldots, \vec\omega_N}$ by
  \begin{align*}
    \abs{\braket{\psi_{B^N}, \psi'_{B^N}}}
    &\geq \sum_{\mathclap{(\vec\omega_1,\ldots,\vec\omega_N) \in \Omega_\varepsilon}} p_{\vec\omega_1,\ldots,\vec\omega_N}
    = 1 - \sum_{\mathclap{(\vec\omega_1,\ldots,\vec\omega_N) \not\in \Omega_\varepsilon}} p_{\vec\omega_1,\ldots,\vec\omega_N} \\
    &= 1 - \PP(\norm{\sum_{n=1}^N X_n - \mathbb E[\sum_{n=1}^N X_n]}_2 \geq \varepsilon) \\
    &\geq 1 - \frac {\mathbb E[\norm{\sum_{n=1}^N X_n - \mathbb E[\sum_{n=1}^N X_n]}_2^2]} {\varepsilon^2} \\
    &= 1 - \frac {\sum_{n=1}^N \mathbb E[\norm{X_n - \mathbb E[X_n]}_2^2]} {\varepsilon^2} \\
    &\geq 1 - \frac {\sum_{n=1}^N \mathbb E[\norm{X_n}_2^2]} {\varepsilon^2},
  \end{align*}
  where we have used Chebyshev's inequality and independence of the $(X_n)$.
  Now, each random variable $X_n$ takes values in the finite set of $\{\vec\omega\}$, so that $\mathbb E[\norm{X_n}_2^2] \leq \max_{\vec\omega} \norm{\vec\omega}_2^2 =: M^2$.
  It follows that
  \begin{equation*}
    \abs{\braket{\psi_{B^N}, \psi'_{B^N}}}
    \geq 1 - \frac {N M^2} {\varepsilon^2}.
  \end{equation*}
  By the bounds relating the fidelity and trace distance applied to $\rho_{B^N} = \proj{\psi}_{B^N}$ and $\rho'_{B^N} = \proj{\psi'}_{B^N}$,
  \begin{equation*}
    \frac 1 2 \norm{\rho_{B^N}- \rho'_{B^N}}_1
    \leq \sqrt{1 - \abs{\braket{\psi_{B^N}, \psi'_{B^N}}}^2}
    \leq
     \sqrt{2 \frac {N M^2} {\varepsilon^2}} =: f(\varepsilon),
  \end{equation*}
  so that \autoref{smoothing tests} gives
  \begin{equation*}
    \beta_\alpha(\rho_{B^N}, \sigma_{B^N}) \geq \beta_{\alpha - f(\varepsilon)}(\rho'_{B^N}, \sigma_{B^N}),
  \end{equation*}
  for all $\sigma_{B^N}$.

  On the other hand, note that we can certainly write $\ket{\psi'_{B^N}}$ as a sum of $(2 \varepsilon + 1)^k$ joint eigenvectors of the $T$-action on $\mathcal H_{B^N}$, as the remaining weights $\sum_n \vec\omega_n$ are integral vectors in $\ZZ^k$ that are all within $\pm \varepsilon$ of the mean $\vec m$. Since the effect of averaging under the group action amounts to a projective measurement in this eigenbasis (dephasing!), \autoref{projective measurement lemma} implies that
  \begin{equation*}
    \beta_{\alpha - f(\varepsilon)}(\rho'_{B^N}, \tilde{\rho}'_{B^N}) \geq \frac {\alpha - f(\varepsilon)} {(2 \varepsilon + 1)^k},
  \end{equation*}
  where $\tilde\rho'$ denotes the $T$-average of $\rho'$.

  Finally, choose $\varepsilon = {\sqrt {8 N} M} / \alpha$ so that $f(\varepsilon) = \alpha/2$,
  and conclude that
  \begin{equation*}
    \beta_\alpha(\rho_{B^N}, \tilde\rho'_{B^N})
    \geq \frac \alpha {2 (2 \varepsilon + 1 )^k}
    \geq \frac {\alpha^{k+1}} {2 (2 {\sqrt {8 N} M} + 1 )^k}
    \geq D \frac {\alpha^{k+1}} {N^{k/2}}
  \end{equation*}
  for some constant $D > 0$ independent of $\alpha$.

  If $\rho_{B^N}$ is an arbitrary separable state then we can still write it as a convex combination $\rho_{B^N} = \sum_i p_i \rho^{(i)}_{B^N}$
  of pure product states. Thus it follows from \eqref{eq:invariant lower bound infsup} and the above that
  \begin{equation*}
    \sup_{\tilde\sigma_{B^N}} \beta_\alpha(\rho_{B^N}, \tilde\sigma_{B^N})
    \geq
    \min_i \sup_{\tilde\sigma_{B^N}} \beta_\alpha(\rho^{(i)}_{B^N}, \tilde\sigma_{B^N})
    \geq
    D \frac {\alpha^{k+1}} {N^{k/2}},
  \end{equation*}
  since any $E_B$ that is feasible for $\rho_{B^N}$ in \eqref{eq:invariant lower bound infsup} is also feasible for at least one of the $\rho^{(i)}_{B^N}$.
\end{IEEEproof}

\begin{cor}[Shot-noise limit]
\label{covariant abelian shot-noise bound}
  Let $G = T$ be abelian.
  For any region estimator for a covariant family of separable probe states on $\mathcal H_B^{\otimes N}$ and uniform prior, we have a lower bound
  \begin{equation}
    \label{eq:invariant abelian shot-noise bound}
    \frac {V_{\max}} {\abs G} \geq D \, p_{\text{succ}}^{\dim T+1} {N^{-\dim T/2}},
  \end{equation}
  where the constant $D > 0$ depends only on the representation $\mathcal H_B$.
  More generally, for an arbitrary prior $p_G$ and all $\alpha < p_\text{succ}^2/2$, we have that
  \begin{equation}
    \label{eq:covariant abelian shot-noise bound}
    \frac {V_{\max}} {\abs G} \geq D \, \alpha^{\dim T} \, \beta_{p_\text{succ}-{\sqrt{2 \alpha}}}(p_G, \frac {\id_G} {\abs G}) \, N^{-\dim T/2}.
  \end{equation}
\end{cor}

Just as \eqref{eq:state-independent covariant heisenberg bound} could be interpreted as a Heisenberg limit, the lower bound \eqref{eq:covariant abelian shot-noise bound} can be understood as a shot-noise bound in each component of the unknown parameter $g \in G$. It would be interesting to generalize \eqref{eq:covariant abelian shot-noise bound} to non-abelian groups $G$.

\medskip\subsubsection*{Stein's lemma}
In view of the generalized Stein's lemma proved in \cite{brandaoplenio10}, one might expect that any lower bound on $\sup_{\tilde\sigma_{B^N}}(\rho_B^{\otimes N}, \tilde\sigma_{B^N})$ should decay exponentially in $N$ rather than polynomially with $N$. However, while this Stein's lemma is indeed applicable to the situation at hand (with alternative hypothesis the set of $G$-invariant states), it can be shown that the corresponding error exponent, which has been called the \emph{$G$-asymmetry} or \emph{regularized relative entropy of frameness} \cite{vaccaro_et_al_2008,gourmarvianspekkens09}, is always equal to zero -- thus there is no contradiction. This has been proved in \cite[Corollary 11]{gourmarvianspekkens09},
but also follows from \autoref{state-independent covariant heisenberg bound} (which furthermore gives more precise information than the subexponential decay predicted by Stein's lemma).

\section{Examples}
\label{sec:example}

\subsection{Phase Estimation}
\label{subsec:phase-estimation}

In this section we apply our bounds to various scenarios for the estimation of a phase parameter in $U(1) = [0,2\pi]/\!\!\sim$.
We shall always equip $U(1)$ with the usual metric and Lebesgue measure of total volume $2\pi$.

\medskip\subsubsection*{Single-Body Hamiltonian}
We first consider a probe system $\mathcal H_{B^N} = \mathcal H_B^{\otimes N}$ and a family of probe states that is obtained by the evolution of an initial state $\rho_B^0$ under the action of a single-body Hamiltonian $H_N = \sum_n H^{(n)}$, where $H^{(n)} = \id^{\otimes n-1} \otimes H \otimes \id^{\otimes N-n}$ and $H$ a periodic Hamiltonian on $\mathcal H_B$. Thus,
\begin{equation*}
  \rho_B^\theta
  = e^{\imath \theta H_N} \rho_B^0 e^{-\imath \theta H_N}
  = (e^{\imath \theta H})^{\otimes N} \rho_B^0 (e^{-\imath \theta H})^{\otimes N}
\end{equation*}
for $\theta \in [0,2\pi]$. Mathematically, $H$ generates a representation of $U(1)$ on the single-body Hilbert space $\mathcal H_B$, and the representation generated by $H_N$ on $\mathcal H_{B^N} = \mathcal H_B^{\otimes N}$ is precisely its $N$-fold tensor product. Thus, \autoref{state-independent covariant heisenberg bound} applied to $X = G = U(1)$ gives the following lower bound that holds for any region estimator and prior $p_{U(1)}$:
\begin{equation*}
  \frac {V_{\max}} {2\pi} \geq C \, \frac {\beta_{p_\text{succ}}(p_{U(1)}, \id_{U(1)} / {2\pi})} N
\end{equation*}
The constant $C$ depends only on the generator $H$; see the first proof of \autoref{heisenberg counting lemma} for an explicit bound. In particular, for the uniform prior $p_{U(1)} = {\id_{U(1)}} / {2\pi}$ we find that
\begin{equation}
\label{eq:state-independent invariant lower bound for phase estimation}
  \frac {V_{\max}} {2\pi} \geq C \, \frac {p_\text{succ}} N.
\end{equation}
If we restrict to separable probe states then by \eqref{eq:invariant abelian shot-noise bound} this can be strengthened to give
\begin{equation}
\label{eq:invariant shot-noise lower bound for phase estimation}
  \frac {V_{\max}} {2\pi} \geq D \, \frac {p_\text{succ}^2} {\sqrt N},
\end{equation}
where again the constant $D$ depends only on the generator $H$.

\begin{figure}
\begin{center}
\includegraphics[width=\linewidth]{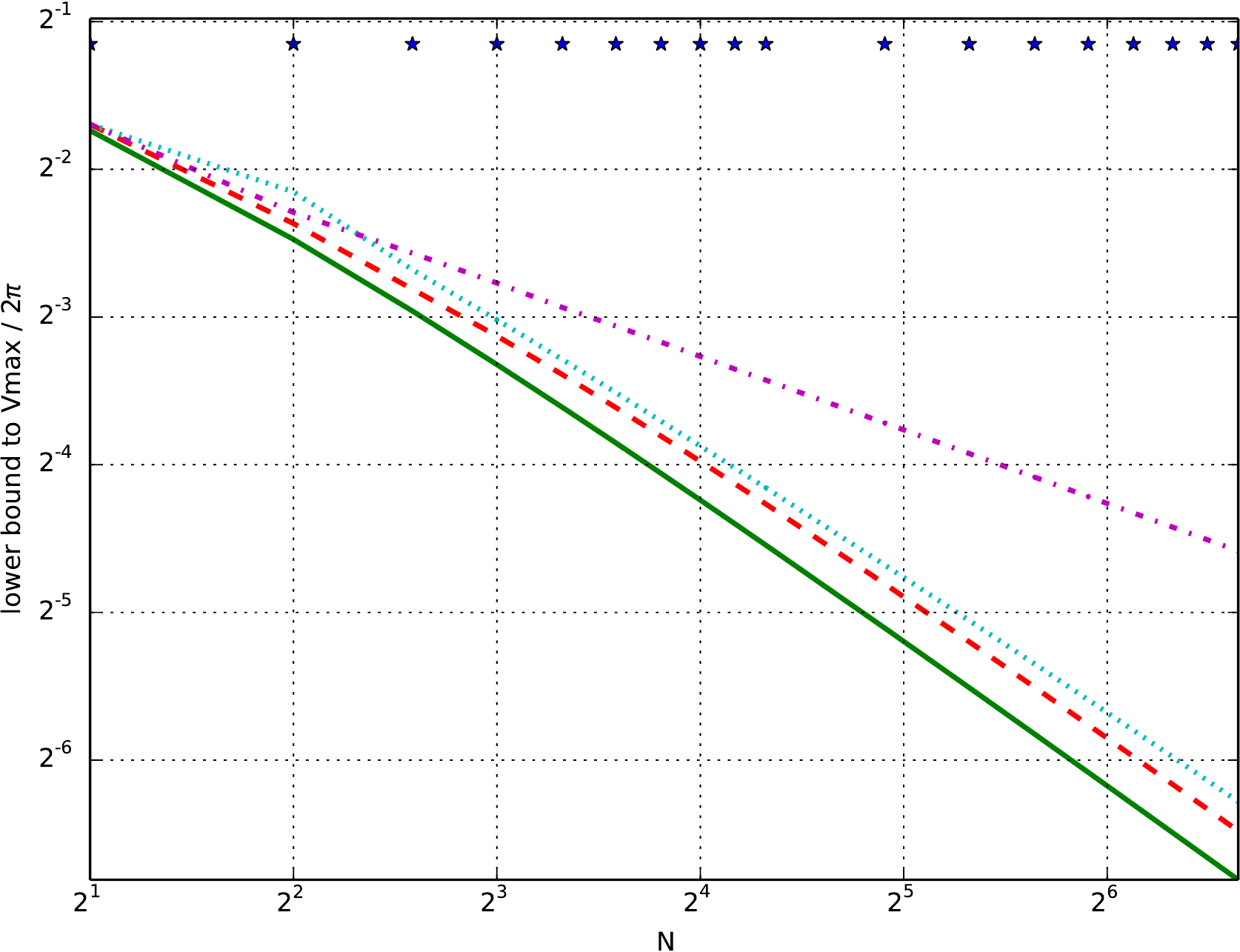}
\end{center}
\caption{\emph{Lower bound \eqref{eq:invariant lower bound supinf} for various probe states, uniform prior and varying $N$ ($\alpha = 0.9$):}
For the states of Sanders--Milburn \cite{sandersmilburn95} (dotted line) and Pegg--Summy/Berry--Wiseman \cite{summy_pegg_1990_phase,berrywiseman00} (dashed line), the lower bound scales asymptotically like the theoretical $1/N$ ``Heisenberg limit'' \eqref{eq:state-independent invariant lower bound for phase estimation} (solid line).
In contrast, tensor powers of the plus state $\ket+ = \frac 1 {\sqrt 2} \left( \ket 0 + \ket 1 \right)$ (dash-dotted line) cannot overcome the $1/{\sqrt N}$ shot-noise limit \eqref{eq:invariant shot-noise lower bound for phase estimation}.
For the GHZ-state $\frac 1 {\sqrt 2} ( \ket 0^{\otimes N} + \ket 1^{\otimes N})$ (stars), though entangled, we get a constant lower bound. Indeed, it is well-known that the GHZ-state achieves Heisenberg scaling only if additional prior information about the phase is available \cite[\S{}III.C]{hall_wiseman_2012}.}
\label{phase estimation numerics figure}
\end{figure}

We can also evaluate the state-dependent lower bound \eqref{eq:invariant lower bound supinf} numerically for individual probe states $\rho^{x_0}_B$, since it is given by a semi-definite program (note that the supremum over $\tilde\sigma_B$ can be incorporated into the dual program). See \autoref{phase estimation numerics figure} for some illustrative numerical results.

We now illustrate how to recover the well-known Heisenberg and shot-noise scaling for the mean-square error $\Delta^2$ of a point estimator.
As explained in \autoref{sec:region estimators}, we can always consider a point estimator as a region estimator by interpreting its estimate as the center of a $\delta$-ball, for any choice of $\delta > 0$. If we do so, then \eqref{eq:maximal volume point estimator}, \eqref{eq:success probability point estimator} and \eqref{eq:state-independent invariant lower bound for phase estimation} combine to
\begin{equation*}
  \frac {2 \delta} {2\pi} \geq C \, \frac {1 - \Delta^2/\delta^2} N.
\end{equation*}
By optimizing over $\delta > 0$, we arrive at the following lower bound, which holds for an arbitrary point estimator and uniform prior:
\begin{equation*}
  \Delta^2 \geq \frac {4 C^2 \pi^2} {27} \frac 1 {N^2}.
\end{equation*}
Thus we have recovered \emph{Heisenberg scaling} for the mean-square error, which is normally established by evaluating the Cr\'{a}mer-Rao bound and averaging over the prior information.
Conversely, since we know that Heisenberg scaling is asymptotically achievable for the mean-square error \cite{summy_pegg_1990_phase,sandersmilburn95,berrywiseman00}, it follows from \eqref{eq:success probability point estimator} that there exist probe states and region estimators for which $V_{\max}$ scales as 
$1 / N$ for any fixed value of $p_\text{succ}$.

Similarly, \eqref{eq:maximal volume point estimator}, \eqref{eq:success probability point estimator} and \eqref{eq:invariant shot-noise lower bound for phase estimation} combine to
\begin{equation*}
  \frac {2 \delta} {2\pi} \geq D \, \frac {(1 - \Delta^2/\delta^2)^2} {\sqrt N} \geq D \, \frac{1 - 2 \Delta^2 / \delta^2} {\sqrt N},
\end{equation*}
which gives the following lower bound that holds for any point estimator, uniform prior and \emph{separable} probe states,
\begin{equation*}
  \Delta^2 \geq \frac {2 D^2 \pi^2} {27} \frac 1 {N}.
\end{equation*}
It exhibits the expected \emph{shot-noise scaling} behavior.
The same procedure can be used to see that the large-$N$ scaling of the curves in \autoref{phase estimation numerics figure} is consistent with the literature.

The above discussions can be readily generalized to the estimation of multiple phases (where $T = U(1)^k$).

\medskip\subsubsection*{Energy-Bounded Probe States}
We now consider phase estimation where the probe system is the single-mode bosonic Fock space $\mathcal H_B = L^2(\RR) \cong \Sym(\CC)$. Let $n = a^\dagger a$ denote the bosonic number operator and $(\ket n)_{n=0}^\infty$ the occupation number basis.
We will establish a fundamental lower for all covariant families of probe states generated by the Hamiltonian $H = a^\dagger a$ with \emph{bounded mean energy} $\braket{\rho^\theta_B, H} = \braket{\rho^0_B, H} \leq E$ \cite{yuenozawa93}.

For this, let $P^{(N)}_B$ denote the projector onto the finite-dimensional subspace $\mathcal H_B^{(N)}$ spanned by $(\ket n)_{n < N}$. Then,
\begin{align*}
  &\quad\braket{\rho^0_B, P^{(N)}_B}
  = 1 - \sum_{n=N}^\infty \braket{n | \rho^0_B | n} \\
  &\geq 1 - \frac 1 N \sum_{n=N}^\infty n \braket{n | \rho^0_B | n}
  \geq 1 - \frac E N,
\end{align*}
and hence the gentle measurement lemma implies that the state $\rho'_B := P_N \rho^0_B P_N / \braket{\rho^0_B,P_N}$ has trace-norm distance at most
\begin{equation*}
  \frac 1 2 \norm{\rho^0_B - \rho'_B}_1 \leq \sqrt{\frac E N}.
\end{equation*}
Observe that $\rho'_B$ is supported on the subspace $\mathcal H_B^{(N)}$, which is $N$-dimensional.
It follows from \autoref{smoothing tests} and \autoref{projective measurement lemma} that
\begin{equation*}
  \beta_\alpha(\rho^0_B, \tilde\sigma_B) \geq
  \beta_{\alpha - \sqrt{E/N}}(\rho'_B, \tilde\sigma_B) \geq
  \frac {\alpha - \sqrt{E/N}} N,
\end{equation*}
where $\tilde\sigma_B$ denotes the $U(1)$-average of $\rho'_B$. Set $N = \lceil 4 E / \alpha^2 \rceil$, so that $\sqrt{E/N} \leq \alpha/2$ and
\begin{equation*}
  \frac {\alpha - \sqrt{E/N}} N
  \geq \frac {\alpha^3} {8E + 2 \alpha^2}.
\end{equation*}
In view of \eqref{eq:invariant lower bound supinf} we thus arrive at the following lower bound, which holds for all region estimators, uniform prior, and probe states whose mean energy is bounded by $E$:
\begin{equation*}
  \frac {V_{\max}} {2 \pi} \geq \frac {p_\text{succ}^3} {8E + 2 p_\text{succ}^2}.
\end{equation*}
which scales as $1/E$ for large $E$.
Heisenberg scaling for the mean-square error, which was established in \cite{giovannettilloydmaccone12}, is an easy consequence of this bound (cf.\ \cite{yuen_1992,nair12,hayashi_2013_fourier}).

\medskip\subsubsection*{Nonlinear Interactions}
The Heisenberg scaling limit can be exceeded in local estimation when the Hamiltonian contains interaction terms. For a Hamiltonian with identical $k$-body interactions between all subsets of $k$ probe systems, the Heisenberg limit for the mean-square error becomes $N^{-k}$~\cite{luis_2004_nonlinear,beltran_luis_2005_breaking,roy_braunstein_2008_exponentially,rey_jiang_lukin_2007_quantumlimited,choi_sundaram_2008_boseeinstein,napolitano_koschorreck_dubost_et_al_2011_interactionbased,boixo_flammia_caves_et_al_2007_generalized}. For instance, for $H_N$ the single-body Hamiltonian from above, a generator of the form $H_N^2$ gives a $1/N^2$ scaling. Additionally, the use of unentangled probe states still gives a scaling of $N^{-k + 1/2}$; the possibility of the Hamiltonian itself generating entanglement closes the gap between the two types of probes.
However, as mentioned in the introduction, such ``super-Heisenberg'' scalings are not to be found in \emph{global} estimation.
Hall and Wiseman offer a resolution of this apparent paradox by analyzing \emph{iterative} schemes in which local estimation is repeatedly employed to perform global estimation~\cite{hall_wiseman_2012}.

A different nonlinear approach was proposed in~\cite{braun_martin_2011_heisenberglimited}, whereby the Hamiltonian couples all $N$ probe systems to an auxiliary system, i.e.\ the total Hamiltonian takes the form $H_N\otimes H'$. Here the goal was to show that separable probe states themselves offer $1/N$ scaling and that the scheme is more resilient to phase noise. Indeed, in~\cite{braun_popescu_2013_coherently} it is claimed that in this scenario a $1/N$ scaling is even possible with (classical) coupled harmonic oscillators.

However, this conclusion does not hold for global estimation. Specifically, we now show that separable probe states have the same performance under a Hamiltonian of the form $H_{N+1} = J_{z,N} \otimes \sigma_z$, with $J_{z,N}=\sum_{n=1}^N\sigma_z^{(n)}$, as they do under $J_{z,N}$ itself. The reasoning is much the same as in the proof of \autoref{prp:separable}. In particular, we need only consider pure product probe states, now of the form  $\ket{\psi_{B^{N+1}}}=\ket{\psi_{B^N}}\otimes \ket{\psi_{B_{N+1}}}$, where $\ket{\psi_{B^N}}=\ket{\psi_{B_1}}\otimes\cdots\otimes \ket{\psi_{B_N}}$. The $\ket{\psi_{B^N}}$ may again be truncated to a state $\ket{\psi'_{B^N}}$ with support on only $O(\sqrt N)$ eigenvectors with weight within $\pm O(\sqrt N)$ of the mean; note that now the weight is just a scalar as there is only one generator. The rest of the proof proceeds exactly as before, though now $M=1$, as again the effect of the group average on the truncated probe state $\rho'_{B^{N+1}}=\proj{\psi_{B^N}'}\otimes\proj{\psi_{B_{N+1}}}$ is to make a projective measurement in the eigenbasis of the generator $H_{N+1} = J_{z,N} \otimes \sigma_z$.
While the latter has $N+1$ distinct eigenvalues, $\rho'$ has support on no more than $O(\sqrt N)$ of them.
Thus we may conclude, as in \autoref{covariant abelian shot-noise bound}, that
\begin{align*}
	\frac {V_{\max}} {2\pi} \geq \frac 1 {O(\sqrt{N})}
\end{align*}
for any prior distribution.

\subsection{State Estimation}
\label{subsec:state estimation}

We now consider the problem of estimating a density matrix $\rho$ with known eigenvalues $r_1,\ldots,r_d$ from $N$ i.i.d\ copies of the state \cite{keyl_2006_quantum} (this is the ``opposite'' of the spectrum estimation problem a la \cite{keylwerner01}).
Thus we consider the parameter space
\begin{equation*}
  X = \{ \rho = \rho^\dagger \in \CC^{d \times d} : \spec \rho = \{r_1, \ldots, r_d\} \},
\end{equation*}
and to each $\rho \in X$, we associate the probe state $\rho_{B^N} = \rho^{\otimes N}$ on $\mathcal H_{B^N} = (\CC^d)^{\otimes N}$.
Observe that $(\rho^{\otimes N})_{\rho \in X}$ is a covariant family with respect to the actions of the non-abelian group $G = \SU(d)$ on $X$ (by conjugation) and on $\mathcal H_{B^N}$ (diagonally).

We shall lower-bound \eqref{eq:invariant lower bound supinf} by evaluating the \emph{dual} formulation \eqref{q hypothesis primal and dual} of $\beta_\alpha(\rho^{\otimes N}, \tilde\sigma)$, with $\tilde\sigma$ the $\SU(d)$-average of $\rho^{\otimes N}$.
For this, recall that $(\CC^d)^{\otimes N}$ is also a representation of the symmetric group $S_N$, which acts by permuting the tensor factors. Since the action of $S_N$ commutes with the one of $\SU(d)$, Schur's lemma implies that the multiplicity spaces of the irreducible $\SU(d)$-representations are representations of $S_N$ (and vice versa).
\emph{Schur--Weyl duality} asserts that these representations are in fact also irreducible and pairwise distinct (e.g.~\cite{goodmanwallach09}). In other words, we have a decomposition of the form
\begin{equation*}
  (\CC^d)^{\otimes N} = \bigoplus_\lambda V_{\SU(d),\lambda} \otimes [\lambda],
\end{equation*}
where $V_{\SU(d),\lambda}$ and $[\lambda]$ are distinct irreducible representations of $\SU(d)$ and $S_N$.
Since $\rho^{\otimes N}$ is permutation-invariant, it follows by Schur's lemma that it is necessarily of the form
\begin{equation*}
  \rho^{\otimes N} = \bigoplus_\lambda p_\lambda \rho_\lambda \otimes \frac {\id_{[\lambda]}} {\dim [\lambda]}
\end{equation*}
Thus the $G$-average of $\rho^{\otimes N}$ is given by
\begin{equation*}
  \tilde\sigma = \bigoplus_\lambda p_\lambda\frac {\id_{V_{SU(d),\lambda}}} {\dim V_{SU(d),\lambda}}  \otimes \frac {\id_{[\lambda]}} {\dim [\lambda]}
\end{equation*}
Let $d_N = \max_\lambda \dim V_{SU(d),\lambda}$, where we take the maximum over all $\lambda$ that occur in the above decomposition. Then
$\rho^{\otimes N} / {d_N} \leq \tilde\sigma,$
i.e.\ $\mu = 1/d_N$, $\tau = 0$ are feasible for the dual formulation \eqref{q hypothesis primal and dual} of $\beta_\alpha(\rho^{\otimes N}, \tilde\sigma)$, and so
$\beta_\alpha(\rho^{\otimes N}, \tilde\sigma) \geq \alpha / {d_N}$.
But recall that we have shown in \autoref{heisenberg counting lemma} that $d_N \leq O(N^{(\dim G - \dim T)/2}) = O(N^{d(d-1)/2})$. Thus we conclude from \eqref{eq:invariant lower bound supinf} that we have the following lower bound, which holds for an arbitrary region estimator and uniform prior:
\begin{equation*}
  \frac {V_{\max}} {\abs X} \geq C \frac {p_\text{succ}} {\sqrt N^{d(d-1)}} = C \frac {p_\text{succ}} {\sqrt N^{\dim X}},
\end{equation*}
where the latter identity holds if we assume that the eigenvalues $r_j$ are all distinct.
Since the probe states $\rho^{\otimes N}$ are all separable, this is the expected shot noise scaling, namely $1/\sqrt N$ in each dimension.

It is interesting to compare this bound with the state-independent ``Heisenberg'' bound \eqref{eq:state-independent invariant lower bound}, where one allows an \emph{arbitrary} covariant family of probe states on $(\CC^d)^{\otimes N}$.
In the non-degenerate case where all $r_j$ are distinct, we have $X \cong G / T$, where $T$ is the subgroup of diagonal matrices. Thus $d_\mu \equiv 1$, and the multiplicities $r^\lambda_\mu$ are the so-called Kostka numbers, which are of the order $O(N^{(d-1)(d-2)/2})$. This can be seen e.g.\ by upper-bounding the Kostka numbers by the Kostant partition function and using \cite[Lemma 1.5]{rassart04}. On the other hand, we recall from \autoref{heisenberg counting lemma} that there are $O(N^{d-1})$ isotypical components of dimension $d_\lambda = O(N^{d(d-1)/2})$ each. Consequently, the bound \eqref{eq:state-independent invariant lower bound} reads
\begin{align*}
  \frac {V_{\max}} {\abs X}
  &\geq D \frac {p_\text{succ}} {N^{(d-1) + d(d-1)/2 + (d-1)(d-2)/2}} \\
  &= D \frac {p_\text{succ}} {N^{d(d-1)}}
  = D \frac {p_\text{succ}} {N^{\dim X}}
\end{align*}
for some constant $D > 0$, corresponding to $1/N$-Heisenberg scaling in each dimension.

\medskip

\begin{figure}
\begin{center}
\includegraphics[width=\linewidth]{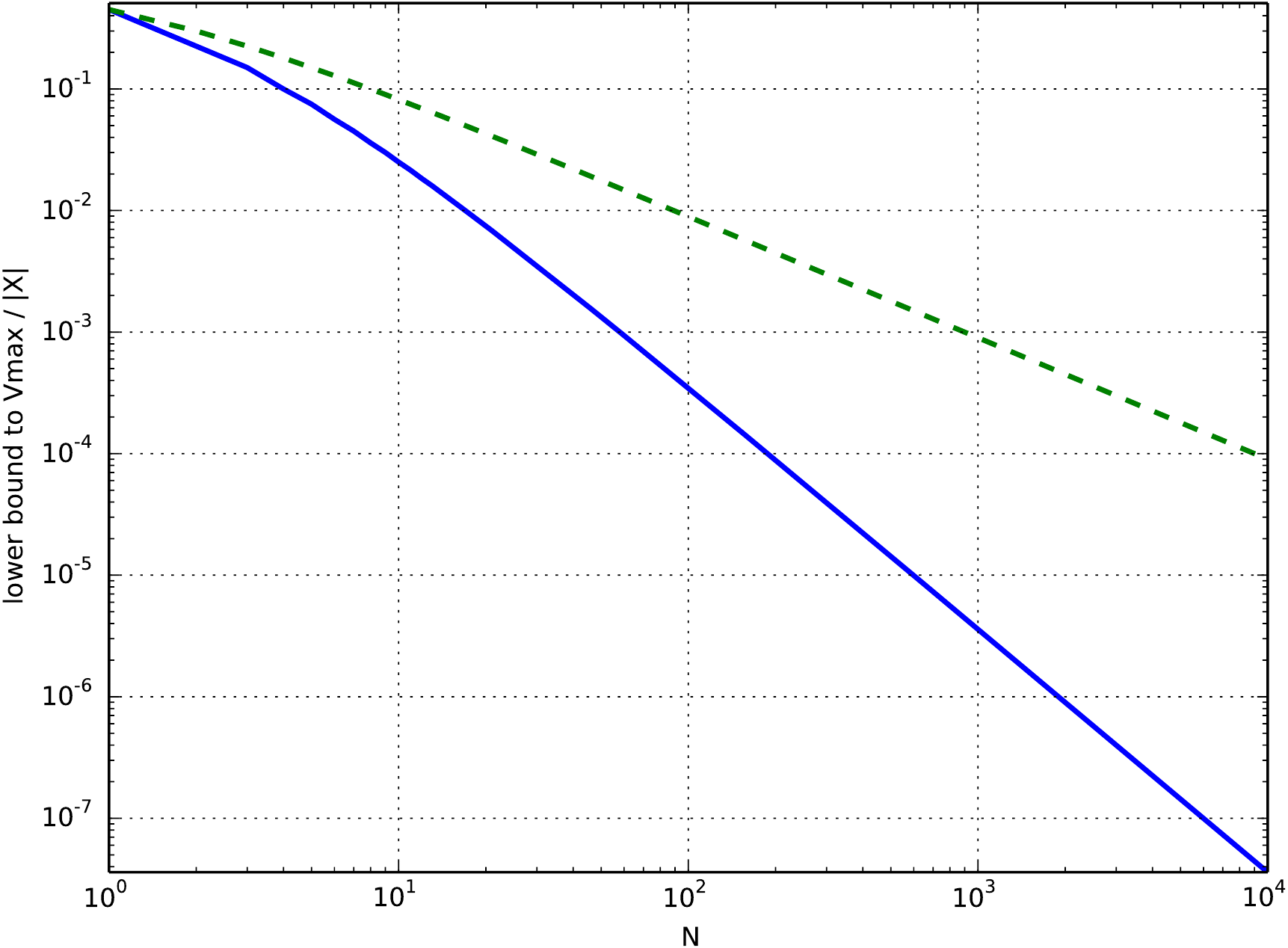}
\end{center}
\caption{\emph{Analytical lower bounds for estimating the pure state of a qubit, $\ket\psi \in \CC^2$, for uniform prior and varying $N$ ($\alpha = 0.9$):}
Given $N$ i.i.d.\ copies of the state, $\ket\psi^{\otimes N}$, we cannot overcome $1/N$-scaling \eqref{eq:pure-state shot-noise scaling} (dashed line).
In contrast, the state-independent lower bound \eqref{eq:pure-state heisenberg scaling},
where we allow for an arbitrary covariant family of probe states on $(\CC^2)^{\otimes N}$,
scales as $1/N^2$ for large $N$ (solid line).
This is consistent with shot-noise resp.\ Heisenberg scaling for the two-dimensional Bloch sphere.}
\label{pure state estimation analytics figure}
\end{figure}

On the other end of the spectrum we may also consider the problem of estimating a \emph{pure state} in $X = \{ \proj \psi : \norm{\psi} = 1 \}$ \cite{hayashi_1998_asymptotic} (corresponding to $r_1 = 1$, $r_2 = \ldots = r_d = 0$). Here, we observe that any i.i.d.\ state $\rho^{\otimes N} = \proj\psi^{\otimes N}$ is completely supported on the symmetric subspace, which is an irreducible representation of $\SU(d)$. Thus the $G$-average is equal to the normalized projector onto the symmetric subspace, whose dimension scales as $O(N^{d-1})$. It follows as above that
\begin{equation}
  \label{eq:pure-state shot-noise scaling}
  \frac {V_{\max}} {\abs X} \geq C \frac {p_\text{succ}} {N^{d-1}} = C \frac {p_\text{succ}} {\sqrt N^{\dim X}}.
\end{equation}
Using the same argument as in \autoref{subsec:phase-estimation}, this implies that the mean-square error of any point estimator scales at best as $1/N$, in agreement with the central limit theorem and \cite[Theorem 4]{hayashi_1998_asymptotic} (where the minimal mean-square error was even computed to third order).

To evaluate the state-independent lower bound \eqref{eq:state-independent invariant lower bound}, where we allow for an arbitrary covariant family on $\mathcal H_{B^N}$, we use that $X \cong \SU(d) / U(d-1)$.
It follows by inspection of the corresponding branching rule that $\sum_\lambda d_\lambda m^\lambda_\mu / d_\mu$ is of the order $O(N^{2(d-1)})$ (\autoref{branching lemma} in \autoref{branching appendix}), so that
\begin{equation}
  \label{eq:pure-state heisenberg scaling}
  \frac {V_{\max}} {\abs X} \geq D \frac {p_\text{succ}} {N^{\dim X}}.
\end{equation}
See \autoref{pure state estimation analytics figure} for an illustration of both bounds in the case of qubits ($d = 2$ and hence $\dim X = 2$).

Thus in both cases we recover the expected shot-noise resp.\ Heisenberg scaling in each dimension of the parameter space, suggesting that it might be possible to generalize the covariant scaling results \autoref{state-independent covariant heisenberg bound} and \autoref{covariant abelian shot-noise bound} to general homogeneous spaces $X$.


\section{Conclusion}

We have derived lower bounds on the sizes of region estimators in quantum parameter estimation. For parameter spaces of finite volume, the bounds can be stated in terms of a hypothesis testing scenario, while for probe states covariant with respect to a group, the bounds can be stated in terms of representation theoretic quantities of the probe systems. The latter bounds are shown to converge to Heisenberg and shot-noise limits for entangled and separable probe states of $N$ systems, in the limit of large $N$.

It would be interesting to relax the assumption that the probe states are covariant with respect to a compact group and, for instance, investigate the minimal sizes of confidence regions when estimating position and momentum shifts of a free particle. Surely the minimal size is of order $\hbar$. Another interesting open question is the extent to which the bounds can be achieved, particularly in the large $N$ limit.
Here one may be able to make general statements, not restricted to particular models, by making use of \emph{local asymptotic normality} results as proposed for point estimators in~\cite{KahnGuta09,YamagataFujiwaraGill13}.

\section*{Acknowledgements}

We thank Mario Berta, Robin Blume-Kohout, Matthias Christandl, Gabriel Durkin, Martin Fraas, David Gross, Volkher Scholz, and Mankei Tsang for valuable discussions and feedback.

\appendices

\section{Measurability}
\label{measurability appendix}

In this section we recall some definitions and facts on the measurability of vector-valued functions that are necessary to rigorously define cq-systems with infinite classical part (see e.g.\ \cite{takesaki79,ryan02} for the general theory).

Let $(X,\mu_X)$ be a 
standard measure space and $\mathcal H_B$ a (not necessarily finite-dimensional) Hilbert space.
Let $B_1(\mathcal H_B)$ denote the space of trace-class operators, i.e.\ operators $A$ for which $\norm{A}_1 := \tr \sqrt{A^\dagger A} < \infty$, and $B(\mathcal H_B)$ denote the space of bounded operators on $\mathcal H_B$ for which the operator norm $\norm{A}_\infty < \infty$. Both spaces are Banach spaces, and the latter is the dual of the former.

A function $\rho_{XB} = (\rho^x_B) \colon X \rightarrow B_1(\mathcal H_B)$ is called \emph{simple} if it is of the form $\rho^x_B = \sum_{j=1}^J A_j \id_{O_j}(x)$ for disjoint measurable subsets $O_j \subseteq X$ and operators $A_j \in B_1(\mathcal H_B)$.
It is called \emph{$\mu_X$-measurable} if there exists a sequence of simple functions that converge $\mu_X$-almost everywhere to $\rho_{XB}$. In particular, this implies that the norm function $x \mapsto \norm{\rho^x_B}_1$ is measurable in the usual sense.

A function $E_{XB} = (E^x_B) \colon X \rightarrow B(\mathcal H_B)$ is called \emph{weak-$\star$-measurable} if the function $x \mapsto \braket{\rho_B, E^x_B}$ is measurable for each $\rho_B \in B_1(H)$.

We now show that the quantities $V_{\max}$ and $p_\text{succ}$ are well-defined by \eqref{eq:success probability} and \eqref{eq:maximal volume}.

\begin{lem}
\label{measurability lemma}
  Let $M_B$ be a POVM with outcomes in some measurable space $Y$, $\mathcal E \subseteq X \times Y$ a measurable subset, and $\rho_{XB} = (\rho^x_B)$ a $\mu_X$-measurable function which takes values in the set of density operators on $\mathcal H_B$.
  \begin{enumerate}
  \item The sets $\mathcal E_x = \{ y : (x,y) \in \mathcal E \}$ and $\mathcal E_y = \{ x : (x,y) \in \mathcal E\}$ are measurable. In particular, $V_{\max}$ is well-defined.
  \item The function $x \mapsto E_B(\mathcal E_x)$ is weak-$\star$-measurable, i.e.\ $\braket{\rho_B, E_B(\mathcal E_x)}$ is measurable for each $\rho_B \in \mathcal S(\mathcal H_B)$.
  \item The function $x \mapsto \braket{\rho^x_B, E_B(\mathcal E_x)}$ is measurable. Therefore, the success probability $p_{\text{succ}}$ is well-defined.
  \end{enumerate}
\end{lem}
\begin{IEEEproof}
  1) This is a standard fact that is routinely established on the route to proving Tonelli's theorems, see e.g.\ \cite[Proposition 2.34, a)]{folland99}.

  2) For each $\rho_B$, $\braket{\rho_B, E_B(-)}$ is an ordinary probability measure on $Y$, so that the measurability of $\braket{\rho_B, E_B(\mathcal E_x)}$ is again standard \cite[Theorem 2.36]{folland99}.

  3) This follows from the $\mu_X$-measurability of $\rho_{XB}$ and point 2 \cite[Lemma 7.15]{takesaki79}.
\end{IEEEproof}

\section{Zero Duality Gap for Binary Hypothesis Testing}
\label{zero duality gap appendix}

We consider binary hypothesis testing between two (normal) states $\rho$ and $\sigma$ in the positive cone $\mathcal M^+_*$ of the predual $\mathcal M_*$ of an arbitrary von Neumann algebra $\mathcal M$ \cite{takesaki79}. In particular, all results in this section are applicable to binary hypothesis testing on classical--quantum systems as defined in \autoref{sec:region estimators} (see below).
We define the norm of a state to be $\abs\rho := \braket{\rho,\id_{\mathcal M}}$.
It will be convenient to require that $\rho$ (but not $\sigma$) be normalized.
A \emph{binary hypothesis test} between $\rho$ and $\sigma$ is a $\{\rho,\sigma\}$-valued measurement, or, equivalently, an observable $f \in \mathcal M$ with $0 \leq f \leq \id_{\mathcal M}$ (the POVM element corresponding to outcome $\rho$). The type I error is $1-\alpha = \braket{\rho, \id_{\mathcal M} - f}$, while the type II error is $\beta = \braket{\sigma, f}$. The \emph{hypothesis testing region} is defined as
\begin{equation*}
  \mathcal R(\rho, \sigma) = \{ (\braket{\rho, f}, \braket{\sigma, f}) : 0 \leq f \leq \id_{\mathcal M} \}.
\end{equation*}
Just as in the classical case \cite[pp.\ 62]{lehmannromano05} we can establish the following properties:

\begin{lem}
  The hypothesis testing region $\mathcal R(\rho, \sigma)$ is a compact, convex subset of the square $[0,1] \times [0,\abs\sigma]$. It contains the diagonal and it is closed under the symmetry $(\alpha,\beta) \mapsto (1-\alpha,\abs\sigma-\beta)$.
\end{lem}
\begin{IEEEproof}
  All assertions except for compactness are immediate.
  To see compactness, observe that the set of POVM elements
  \begin{equation*}
    \{ f \in \mathcal M : 0 \leq M \leq \id_{\mathcal M} \}
    = B_1(\mathcal M) \cap \mathcal M_+
  \end{equation*}
  is a weak-$\star$-closed subset of the unit ball, and therefore also weak-$\star$-compact by the Banach--Alaoglu theorem
  \cite[Theorem 4.3]{rudin_1991}.
  Since $(\braket{\rho,-}, \braket{\sigma,-})$ is weak-$\star$-continuous (by definition), the assertion follows.
\end{IEEEproof}

As a consequence, the infimum in the \emph{lower boundary function}
\begin{align*}
  \beta_\alpha(\rho, \sigma)
  = &\inf \{ \beta : (\alpha, \beta) \in \mathcal R(\rho, \sigma) \} \\
  = &\inf \{ \braket{\sigma, f} : 0 \leq f \leq \id_{\mathcal M}, \braket{\rho, f} = \alpha \} \\
  = &\inf \{ \braket{\sigma, f} : 0 \leq f \leq \id_{\mathcal M}, \braket{\rho, f} \geq \alpha \}
\end{align*}
is attained, and $\beta_\alpha$ is a continuous, convex function of $\alpha \in [0,1]$ \cite[Theorem 10.2]{rockafellar72}.

Note that $\beta_\alpha(\rho, \sigma)$ can be understood as a linear cone program in $\mathcal M$. We define the \emph{(pre)dual program} by
\begin{align*}
  &\beta^*_\alpha(\rho, \sigma) \\
  := &\sup \{ \alpha \mu - \braket{\tau, \id_\mathcal M} :
  \mu \in \RR, \tau \in \mathcal M_*,
  \tau \geq 0,
  \tau \geq \mu \rho - \sigma \} \\
  = &\sup \{ \alpha \mu - \braket{\tau, \id_\mathcal M} :
  \mu \geq 0, \tau \in \mathcal M_*,
  \tau \geq 0,
  \tau \geq \mu \rho - \sigma \}.
\end{align*}
Again, it holds as in the classical case \cite{krafftwitting66} that there is zero duality gap. We adapt an argument sketched in \cite{polyanskiy13}:

\begin{prp}
  We have $\beta_\alpha(\rho, \sigma) = \beta_\alpha^*(\rho, \sigma)$.
\end{prp}
\begin{IEEEproof}
  Let us denote by $f^L$ the conjugate, or Legendre(--Fenchel) transform, of a function $f$.
  Since $\beta_\mu$ is a closed proper convex function, it is equal to its double Legendre transform \cite[Corollary 12.2.1]{rockafellar72}:
  \begin{equation}
  \label{legendre}
    \beta_\alpha(\rho, \sigma) =
    \beta_\alpha^{LL}(\rho, \sigma) =
    \sup \{ \alpha \mu - \beta_\mu^L(\rho, \sigma) : \mu \in \RR \}.
  \end{equation}
  We compute:
  \begin{align*}
    \beta_\mu^L(\rho, \sigma)
    = &\sup \{ \mu \alpha - \beta_\alpha(\rho, \sigma) : \alpha \in [0,1] \} \\
    = &\sup \{ \mu \braket{\rho, f} - \braket{\sigma, f} : 0 \leq f \leq \id_{\mathcal M} \} \\
    = &\sup \{ \braket{\mu\rho - \sigma, f} : 0 \leq f \leq \id_{\mathcal M} \}.
  \end{align*}
  The supremum is attained when $f$ is the projection onto the positive part $\{\mu\rho - \sigma\}_+$ of $\mu\rho - \sigma$ (which exists, since $\mathcal M$ is a von Neumann algebra).
  Thus the Legendre transform of $\beta_\mu$ is given by
  \begin{equation*}
    \beta_\mu^L(\rho, \sigma) = \braket{\{\mu\rho - \sigma\}_+, \id_{\mathcal M}}.
  \end{equation*}
  The claim follows by plugging this into the right-hand side of \eqref{legendre} and comparing it with the definition of $\beta^*_\alpha(\rho, \sigma)$.
\end{IEEEproof}

Let $D(\rho, \sigma)$ denote the relative entropy between two states $\rho$ and $\sigma$ in $\mathcal M_*$ \cite[\S 5]{ohya_petz_1993_quantum}. Then we have the following relation between the relative entropy and the hypothesis-testing region,
\begin{equation}
\begin{aligned}
  \label{eq:hypothesis test dpi}
  &\quad D(\rho \Vert \sigma) \\
  &\geq D(\{\alpha,1-\alpha\} \Vert\{\beta,\abs\sigma-\beta\}) \\
  &= d(\alpha \Vert \beta / \abs \sigma) - \log \abs\sigma
  \quad \forall (\alpha,\beta) \in \mathcal R(\rho,\sigma)
\end{aligned}
\end{equation}
where $d(p \Vert q) := p \log \frac p q + (1-p) \log \frac {1-p} {1-q}$ denotes the binary relative entropy function.
It is an immediate consequence of the data-processing inequality \cite[Corollary 5.12, (iii)]{ohya_petz_1993_quantum}, applied to the channels that correspond to tests (cf.\ \cite[(154)]{polyanskiy_poor_verdu_2010_channel}).
In particular, \eqref{eq:hypothesis test dpi} holds for $\beta = \beta_\alpha(\rho, \sigma)$ (use the optimal test).

\medskip

Now consider cq-hypothesis testing as discussed in \autoref{sec:region estimators}.
We define $\mathcal M_{XB} = L^\infty(X, \mu_X; B(\mathcal H_B))$ as the space of all weak-$\star$-measurable functions for which $\norm{E_{XB}}_\infty := \sup_x \norm{E^x_B}_\infty < \infty$, identifying any two functions that are $\mu_X$-almost everywhere equal.
Then $\mathcal M_{XB}$ is a von Neumann algebra.
Its predual is given by $(\mathcal M_{XB})_* = L^1(X, \mu_X; B_1(\mathcal H_B))$, the space of all $\mu_X$-measurable functions for which $\norm{\rho_{XB}}_1 := \int_X d\mu_X(x) \norm{\rho^x_B}_1 < \infty$ \cite[Theorem IV.7.17]{takesaki79}.
Thus the set of unnormalized cq-states as defined in \autoref{sec:region estimators} agrees precisely with the positive cone $(\mathcal M_{XB})_*^+$ of the predual, and it is easily seen that the definitions of $\beta_\alpha$ and $\beta_\alpha^*$ for general von Neumann algebras reduce to the definitions given in \autoref{sec:region estimators} for cq-systems.

\section{A Technical Lemma}
\label{branching appendix}

\begin{lem}
  \label{branching lemma}
  Let $\mathcal H_{B^N} = (\CC^d)^{\otimes N}$, $G = U(d)$ acting diagonally by $U^{\otimes N}$, and $K = U(d-1) \subseteq G$ embedded as the upper-left block. Then we have that
  \begin{equation*}
    \max_\mu \sum_\lambda d_\lambda r^\lambda_\mu / d_\mu = O(N^{2(d-1)}),
  \end{equation*}
  where $d_\lambda$, $d_\mu$ and $r^\lambda_\mu$ are defined as in \autoref{state-independent invariant lower bound}.
\end{lem}
\begin{IEEEproof}
  We can label the irreducible representations of $U(d)$ by \emph{Young diagrams}, i.e.\ non-increasing sequences $\lambda = (\lambda_1, \ldots, \lambda_d)$ of non-negative integers. By the Weyl dimension formula \eqref{eq:weyl dimension formula}, the dimension of the corresponding irreducible representation $V_{U(d),\lambda}$ is then given by
  \begin{equation}
  \label{eq:weyl dimension SU}
    d_\lambda = \prod_{1 \leq i < j \leq d} \frac {\lambda_i - \lambda_j + j - i} {j - i}.
  \end{equation}

  From Schur--Weyl duality, it is well-known that the irreducible representations $V_{U(d),\lambda}$ that appear in $\mathcal H_{B^N}$ are precisely those that correspond to Young diagrams with $\abs\lambda := \sum_j \lambda_j = N$. The restriction of such an irreducible representation to the subgroup $K = U(d-1)$ is given by the following branching rule \cite{weyl_1946_classical},
  \begin{equation*}
    V_{U(d),\lambda} \big|^{U(d)}_{U(d-1)} = \bigoplus_\mu V_{U(d-1),\mu},
  \end{equation*}
  where the direct sum runs over all Young diagrams $\mu = (\mu_1, \ldots, \mu_{d-1})$ for which
  \begin{equation}
    \label{eq:branching rule weyl}
    \lambda_j \geq \mu_j \geq \lambda_{j+1}.
  \end{equation}
  Note that all (non-zero) multiplicities $m^\lambda_\mu$ are equal to one.
  Since $r^\lambda_\mu \leq m^\lambda_\mu$, and using the dimension formula \eqref{eq:weyl dimension SU}, it follows that
  \begin{equation*}
    d_\lambda r^\lambda_\mu / d_\mu
    = d_\lambda / d_\mu
    = \frac 1 {d!}
      \frac
      {\prod_{1 \leq i < j \leq d} (\lambda_i - \lambda_j + j - i)}
      {\prod_{1 \leq i < j \leq d-1} (\mu_i - \mu_j + j - i)}.
  \end{equation*}
  By the branching rule \eqref{eq:branching rule weyl}, we have that $\mu_i - \mu_j \geq \lambda_{i+1} - \lambda_j$, and so $\mu_i - \mu_j + j - i \geq \lambda_{i+1} - \lambda_j + j - (i+1)$, which is non-trivial if $i + 1 < j$.
  We can thus upper-bound the above expression by
  \begin{align*}
    &\frac 1 {d!}
    \frac
    {
    \prod_{1 < j \leq d-1} (\lambda_1 - \lambda_j + j - 1)
    \prod_{1 \leq i < d} (\lambda_i - \lambda_d + d - i)
    }
    {\prod_{1 \leq i < d-1} (\mu_i - \mu_{i+1} + 1)} \\
    \leq
    &\frac
    {O(N^{d-2} N^{d-1})}
    {\prod_{1 \leq i < d-1} (\mu_i - \mu_{i+1} + 1)},
  \end{align*}
  since each $\lambda_i - \lambda_j \leq \lambda_i \leq N$.
  On the other hand, observe that the branching rule \eqref{eq:branching rule weyl} also implies that $\mu_i \geq \lambda_{i+1} \geq \mu_{i+1}$ for all $\lambda$ that are compatible with a given $\mu$. Since the sum of all $\lambda_j$ has to be equal to $N$, we find that the number of compatible $\lambda$ can be upper-bounded by
  \begin{equation*}
    N \prod_{1 \leq i < d-1} (\mu_i - \mu_{i+1} + 1).
  \end{equation*}
  We conclude that
  \begin{equation*}
    \sum_\lambda d_\lambda r^\lambda_\mu / d_\mu \leq O(N^{2(d-1)}).
  \end{equation*}
\end{IEEEproof}

\bibliographystyle{IEEEtran}
\bibliography{IEEEabrv,lowerbounds}

\begin{IEEEbiographynophoto}{Michael Walter}
first studied mathematics and computer science at the University of Karlsruhe; later, he moved to the University of G\"ottingen, where he obtained a Diplom in mathematics in 2010.
He then moved to ETH Zurich to pursue graduate studies in theoretical physics, obtaining a Ph.D.\ in physics in 2014. 
Currently he is a postdoctoral scholar in physics at Stanford University.
\end{IEEEbiographynophoto}

\begin{IEEEbiographynophoto}{Joseph M.~Renes}
received a B.S.~degree in physics from Caltech in 1999.
Subsequently he moved to the University of New Mexico to study quantum information theory, where he received a Ph.D.\ in physics in 2004.

Between 2005 and 2007 he was an Alexander von Humboldt research fellow, first at the University of Erlangen, and later at the Technical University of Darmstadt.
Currently he is a senior scientist at ETH Zurich.

Dr.~Renes is a member of the American and German Physical Societies, in addition to the IEEE.
\end{IEEEbiographynophoto}

\end{document}